\documentstyle[prd,aps,epsfig,floats,axodraw]{revtex}
\begin{document}
\draft

\renewcommand{\topfraction}{0.8} 
\newcommand{\beq}{\begin{equation}}
\newcommand{\eeq}{\end{equation}}
\newcommand{\bea}{\begin{eqnarray}}
\newcommand{\eea}{\end{eqnarray}}
\newcommand{\pbar}{\not{\!\partial}}
\newcommand{\dbar}{\not{\!{\!D}}}
\def\lsim{\:\raisebox{-0.75ex}{$\stackrel{\textstyle<}{\sim}$}\:}
\def\gsim{\:\raisebox{-0.75ex}{$\stackrel{\textstyle>}{\sim}$}\:}
\twocolumn[\hsize\textwidth\columnwidth\hsize\csname 
@twocolumnfalse\endcsname

\title{Thermalization after inflation and production of massive stable
particles}   
\author{Rouzbeh Allahverdi and Manuel Drees~}     
\address{Physik Department, TU M\"unchen, James Frank
Strasse, D-85748, Garching, 
Germany.}
\date{\today} 
\maketitle
\begin{abstract}
We discuss thermalization through perturbative inflaton decay at the
end of inflation. We find that a thermal plasma should form well
before all inflatons have decayed, unless all gauge symmetries are
badly broken during that epoch. However, before they thermalize, the
very energetic inflaton decay products can contribute to the
production of massive stable particles, either through collisions with
the thermal plasma, or through collisions with each other. If such
reactions exist, the same massive particles can also be produced
directly in inflaton decay, once higher--order processes are
included. We show that these new, non--thermal production mechanisms
often significantly strengthen constraints on the parameters of models
containing massive stable particles; for example, stable charged
particles with mass below the inflaton mass seem to be essentially
excluded.    
\end{abstract}

\pacs{PACS numbers:90.80.Cq, 11.30.Pb \hspace*{1.3cm}
TUM--HEP--461/02} 

\vskip2pc]

\section{Introduction}

According to inflationary models \cite{infl}, which were first
considered to address the flatness, isotropy, and (depending on the
particle physics model of the early Universe) monopole problems of the
hot Big Bang model, the Universe has undergone several stages during
its evolution. During inflation, the energy density of the Universe is
dominated by the potential energy of the inflaton and the Universe
experiences a period of superluminal expansion. After inflation,
coherent oscillations of the inflaton dominate the energy density of
the Universe.
At some later time these coherent oscillations decay to the fields to
which they are coupled, and their energy density is transferred to
relativistic particles; this reheating stage results in a
radiation--dominated Friedmann--Robertson--Walker (FRW) universe.
After the inflaton decay products thermalize, the dynamics of the
Universe will be that of the hot Big Bang model.

Until a few years ago, reheating was treated as the perturbative, one
particle decay of the inflaton with decay rate $\Gamma_{\rm d}$
(depending on the microphysics), leading to the simple estimate
$T_{\rm R} \sim {({\Gamma}_{\rm d}{M}_{\rm Planck})}^{1/2}$ for the
reheat temperature \cite{reheat}\footnote{From now on $M_{\rm Planck}
= 2.4 \times 10^{18}$ GeV represents the reduced Planck mass.}. The
reheat temperature should be low enough so that the GUT symmetry is
not restored and the original monopole problem is avoided. In many
supersymmetric models there are even stricter bounds on the reheat
temperature. Gravitinos (the spin$-3/2$ superpartners of gravitons)
with a mass in the range of 100 GeV to 1 TeV (as expected for
``visible--sector'' superparticles) decay during or after Big Bang
nucleosynthesis. They are also produced in a thermal bath,
predominantly through $2 \rightarrow 2$ scatterings of gauge and
gaugino quanta \cite{gravitino}. This results in the
bound\footnote{This bound does not hold in models with gauge-mediated
supersymmetry breaking, where the gravitino is much lighter than
sparticles in the visible sector, so that the decay of the lightest
visible sparticle into the gravitino occurs well before
nucleosynthesis; nor does it hold in models with anomaly-mediated
supersymmetry breaking, where the gravitino mass exceeds
visible--sector sparticle masses by an inverse loop factor, so that
gravitino decays are sufficiently rapid. We will comment on this
second scenario later.} ${T}_{\rm R} \leq {10}^{7}-{10}^{9}$ GeV, in
order to avoid gravitino overproduction which would destroy the
successful predictions of nucleosynthesis by its late decay
\cite{bbn}.

It has been noticed in recent years that the initial stage of inflaton
decay might occur through a complicated and non--perturbative process
called parametric resonance\footnote{Recently, non-thermal production
of helicity $\pm 3/2$ gravitino \cite{maroto}, and helicity $\pm 1/2$
gravitino \cite{kallosh} from inflaton oscillations have been
considered. For models with a single chiral supermultiplet, the
helicity $\pm 1/2$ component of the gravitino is the superpartner of
the inflaton known as inflatino. The decay channels of the inflatino
have been discussed in Ref.~\cite{rouzbeh}. Also, it has been
suggested \cite{rouzbeh}, and explicitly shown \cite{nps} that in
realistic models with several chiral multiplets helicity $\pm 1/2$
gravitino production is not a problem, so long as the inflationary
scale is sufficiently higher than the scale of supersymmetry breaking
in the hidden sector and the two sectors are coupled only
gravitationally.}, leading to an out--of--equilibrium distribution of
final state particles with energies considerably higher than the
inflaton mass (the preheating stage) \cite{preheat1,preheat2}.
Parametric resonance is particularly efficient for decays to bosonic
degrees of freedom. For inflaton, $\phi$, coupling to scalars $\chi$
via four-leg interactions $h^2 \phi^2 \chi^2$ or three-leg
interactions $h m_\phi \phi \chi^2$ parametric resonance occurs if the
coupling $h$ satisfies \cite{preheat2}
\beq \label{hbound}
h \bar \phi > m_\phi,
\eeq
where $\bar \phi$ is the amplitude of the oscillations of the inflaton
field, and $m_\phi$ is the frequency of oscillations, which is equal
to the inflaton mass during oscillations\footnote{In this article we
assume that $m_\phi$ is essentially constant between $t_{\rm I}$ and
$t_{\rm d}$.}. The explosive decay of the inflaton oscillations via
such couplings comes as a result of two effects. $\chi$ quanta with
physical momentum $k \lsim \sqrt{h \bar \phi m_\phi}$ are produced
during a very short interval each time that $\phi$ passes through the
origin. These quanta subsequently result in an even faster particle
production once their occupation number exceeds unity. The combined
effect destroys the coherent oscillations of the inflaton very rapidly
and brings degrees of freedom to which are coupled to the inflaton to
local equilibrium with it and among themselves \cite{fk}. After some
time the momentum of particles in this $\it{preheat~plasma}$ is
redshifted below the inflaton mass and another stage of inflaton decay
will occur. The nature of inflaton couplings determines what exactly
happens at this stage. It is known that four-leg interactions $h^2
\phi^2 \chi^2$ alone do not result in complete decay of the inflaton
and hence interactions which are linear in the inflaton field will be
required. If such interactions have a large coupling, so that
(\ref{hbound}) is satisfied, the final stage of the inflaton decay
will be very quick. This usually results in a high reheat temperature
which could be problematic in supersymmetric models. Moreover, the
bulk of the energy density may remain in coherent oscillations of the
inflaton until the final decaying stage even if the inflaton coupling
satisfies the condition in (\ref{hbound}). This happens for resonant
decay to fermions \cite{fermions} where decay products can not attain
occupation numbers larger than one. It can also be the case for
bosonic parametric resonance once one goes beyond the simplest toy
models. For example, in the $\it{instant~ preheating}$ scenario
\cite{instant} the $\chi$ quanta may not build up a large occupation
number, if they quickly decay to other fields after each interval of
production. Also, moderate self-interactions of final state bosons
renders resonant decay much less efficient \cite{ac,pr}.

It is therefore generally believed that an epoch of (perturbative)
reheating from the decay of massive particles (or coherent field
oscillations, which amounts to the same thing) is an essential
ingredient of any potentially realistic cosmological model
\cite{jed}. In what follows we generically call the decaying particle
the ``inflaton'', since we are (almost) certain that inflatons indeed
exist. Note also that in a large class of well-motivated models, where
the inflaton resides in a ``hidden sector'' of a supergravity theory
\cite{rs}, its couplings are suppressed by inverse powers of $M_{\rm
Planck}$, and hence so weak that inflaton decays are purely
perturbative. However, it should be clear that our results are equally
well applicable to any other particle whose (late) decay results in
entropy production.

Inflaton decays can only be described by
perturbation theory in a trivial vacuum if the density $n$ of
particles produced from inflaton decay is less than the particle
density in a thermal plasma with the same energy density $\rho$, i.e.
\beq \label{nbound}
n^{1/3} < \rho^{1/4}.
\eeq

Even before all inflatons decay, the decay products form a plasma
which, upon a very quick thermalization, has the instantaneous
temperature 
\beq \label{temp}
T \sim \left( g_*^{-1/2} H \Gamma_{\rm d} M^{2}_{\rm Planck}
\right)^{1/4}, 
\eeq
where $H$ is the Hubble parameter and $g_*$ denotes the number of
relativistic degrees of freedom in the plasma\footnote{We assume that
this number remains essentially constant for all $T > T_{\rm
R}$.}. This temperature reaches its maximum $T_{\rm max}$ soon after
the inflaton field starts to oscillate around the minimum of its
potential, which happens for a Hubble parameter $H_I \leq m_\phi$. If
inflaton decays at this early time are to be described by perturbation
theory in a trivial vacuum, the constraint (\ref{nbound}) should be
satisfied already at $H_I$.  This implies $T_{\rm max} < m_\phi$,
i.e. $\Gamma_{\rm d} < m^3_\phi / M^2_{\rm Planck}$. Note that in
chaotic inflation, where $\bar \phi \sim M_{\rm Planck}$ initially, the same
bound follows from the constraint (\ref{hbound}), since $\Gamma_{\rm
d} \sim h^2 m_\phi$. We note in passing that requiring (\ref{nbound})
to hold only at $H \sim \Gamma_{\rm d}$, where the bulk of inflaton
decays take place, leads to the considerably weaker constraint
$\Gamma_{\rm d} < m^2_\phi / M_{\rm Planck}$. Indeed, this weaker
constraint is satisfied in all potentially realistic inflationary
models we know of that satisfy (\ref{hbound}). If $m^2_\phi / M_{\rm
Planck} > \Gamma_{\rm d} > m^3_\phi / M^2_{\rm Planck}$, the dynamics
of the Universe for $H \lsim \Gamma_{\rm d}$ will be as in the
standard hot Big Bang scenario. However, for some period of time with
$H < m_\phi$, the thermalization of inflaton decay products would have
to {\em reduce} the number of particles. This indicates that in fact
one--particle decay may not have been the dominant mode for reducing
the number of inflatons during that period. For the remainder of this
article we will assume that $T_{\rm max}$ is indeed smaller than
$m_\phi$.

In addition to the thermalized plasma with temperature given by
eq.(\ref{temp}), there will be inflaton decay products that haven't
been thermalized yet, with energy $\simeq m_\phi/2$\footnote{Here we
consider two body decays of the inflaton. The average energy of decay
products will be of the same order even if the inflaton dominantly
decays to three body, or higher, final states.}. During this era the
energy density of the Universe is still dominated by the
(non--relativistic) inflatons that haven't decayed yet. The scale
factor of the Universe $a$ then varies as $a \propto T^{-8/3}$
\cite{kt}. The Universe remains in this phase as long as $H >
\Gamma_{\rm d}$. This can have various cosmological implications, for
example for Affleck--Dine baryogenesis \cite{ace,ad} and electroweak
baryogenesis \cite{dlr}. Here we discuss the production of massive
long--lived or stable particles \cite{ckr,gkr,ky,rosenfeld}.

Particles with mass $m_\chi \leq T_{\rm R}$, and not too small
coupling to the thermal bath of inflaton decay products, are in
thermal equilibrium and their abundance satisfies $n \sim T^{3}_{\rm
R}$ at $H \simeq \Gamma_{\rm d}$. On the other hand, once all
inflatons have decayed, stable particles with mass $m_\chi > T_{\rm
R}$ can only be pair--produced by the Wien's tail of the thermal
spectrum; one might thus naively expect their abundance to be
suppressed by the Boltzmann factor $\exp(-T_{\rm R}/m_\chi)$. This
need not be true if these particles are produced at $H > \Gamma_{\rm
d}$. Particles with mass $T_{\rm R} < m_\chi <T_{\rm max}$ and not too
small coupling have an abundance $\propto T^3$ at early times. Once
the plasma temperature drops below $m$, pair annihilation to and pair
creation from lighter species keeps the particle abundance at
equilibrium so long as $\Gamma_{\rm ann} \geq H$. Finally, at $T =
T_{\rm f}$, the comoving number density of the particle freezes at its
final value. The physical number density of the particle at $T =
T_{\rm R}$ is redshifted by a factor of ${(T_{\rm R}/T_{\rm
f})}^8$. This suggests that the thermal abundance of particles with
mass $T_{\rm R} < m_\chi < T_{\rm max}$ is in general only power--law
suppressed. This has important consequences for the production of
heavy (i.e. $m_\chi > T_{\rm R}$) stable particles \cite{ckr} in
general, e.g. for the Lightest Supersymmetric Particle (LSP)
\cite{gkr} and charged stable particles \cite{ky} in models with low
reheat temperature.

To date, almost all studies have considered the creation of heavy
particles from the scattering of two particles in the thermal
distribution. On the other hand, particles produced from inflaton
decay have an energy $E \simeq m_\phi/2$ before they thermalize; as
already noted, such very energetic particles will exist until inflaton
decay completes at $H \simeq \Gamma_{\rm d}$. We argued above that for
perturbative inflaton decay $m_\phi > T_{\rm max}$. It is thus
possible that heavy particles are efficiently produced from the
scattering of such a ``hard'' particle (with energy $E \simeq
m_\phi/2$) off ``soft'' particles in the thermal bath (with energy $E
\sim T$), or off another ``hard'' particle. Note that this allows to
produce particles with mass $m_\chi > T_{\rm max}$. Of course, the
hard particles eventually come into equilibrium with the thermal bath;
the competition among different interactions thus determines the
abundance of heavy particles produced through this non--thermal
mechanism. Another possibility is that the heavy particles are
themselves produced directly in inflaton decays \cite{rosenfeld}, in
which case their abundance can be even higher.

In this article we study these issues. We begin by a review of
thermalization of the perturbative inflaton decay products, and verify
that a thermal plasma can indeed build up well before the completion
of inflaton decay, if the inflaton decay products couple to some light
(or massless) gauge bosons. We then turn to particle production from
hard--soft and hard--hard scatterings, and from inflaton decay,
estimating the abundance of produced particles in each case. It is
shown that heavy particles may be produced more abundantly than
previously thought, in particular directly from inflaton decay. We
will finally close with some concluding remarks.

\section{Thermalization after inflation}

After inflation the inflaton field starts to oscillate coherently
around the minimum of its potential; at some later time it will decay
to the fields to which it is coupled. In the perturbative regime the
decay occurs over many oscillations of the inflaton field, since
$\Gamma_{\rm d} \ll m_\phi$. This means that the oscillating inflaton
field behaves like non--relativistic matter consisting of a condensate
of zero--mode bosons with mass $m_\phi$. The decay rate $\Gamma_d$ of
the oscillating inflaton field is then identical to the total decay
rate of free on--shell inflaton quanta \cite{reheat}. Most inflatons
have decayed by
$t_{\rm d} = (2/3) \Gamma^{-1}_{\rm d}$. By that time the bulk of the
energy density in the coherent oscillations of the inflaton field has
thus been transferred to relativistic particles with an initial energy
of order $m_\phi/2$. From then on the Universe is
radiation--dominated, and the energy of relativistic particles is
redshifted as $a^{-1} \propto t^{-1/2}$, where $a$ is the scale factor
in the FRW metric. The thermalization of the inflaton decay products
then sets the stage for the familiar hot Big Bang Universe.

However, inflaton decay does not suddenly happen at $t_{\rm
d}$. Rather, it is a prolonged process which starts once the inflaton
field oscillations commence at $t = t_{\rm I}$, when $H = H_{\rm
I} \sim m_\phi$. The comoving number density of zero--mode inflaton
quanta at time $t$, $n_c(t)$, obeys the relation $n_{\rm c}(t) =
n_{\rm I} \exp[-\Gamma_{\rm d}(t-t_{\rm I})]$, where $n_{\rm I}$ is
the inflaton number density at $t_{\rm I}$. For $t_{\rm I} \leq t \leq
t_{\rm d}$ the Universe is matter--dominated, which implies $a \propto
t^{2/3}$, i.e.  $H^{-1} = (3/2)t$. In the time interval between $t$
and $t + (2/3)H^{-1} = 2t$, decay products with energy $\simeq
m_\phi/2$ are produced and their physical number density $\bar n_h$ at
the
end of this interval is (assuming $t_{\rm I} \ll t_{\rm d}$):
\beq \label{infden}
\bar n_h \simeq n_{\rm I} [\exp{(-\Gamma_{\rm d}t)} -
\exp{(-2\Gamma_{\rm d}t)}] {\left ({H \over 2H_{\rm I}} \right )}^{2},
\eeq
where the last factor comes from the expansion of the physical volume.
Particles which were produced earlier have a redshifted energy and
number density. The spectrum of inflaton decay products has been
derived in \cite{mcdonald,a} and it has been shown that, if the
inflaton decay products do not interact with each other, the number
density and energy density of the plasma is dominated by particles
with energy $\sim m_\phi/2$ in the spectrum.

Thermalization is a process during which the energy density $\rho$ of
a distribution of particles remains constant, while their number
density $n$ changes in such a way that the mean energy $\bar E$ of
particles reaches its equilibrium value $\sim T$. For a distribution
of relativistic particles which consists of $n_{\rm B}$ bosonic and
$n_{\rm F}$ fermionic degrees of freedom in thermal equilibrium, and
with negligible chemical potential, we
have $\rho = \pi^2/30 \left(n_{\rm B} + {7 \over 8}n_{\rm F}\right) T^4$
and $n = \zeta(3)/\pi^2 \left(n_{\rm B} + {3 \over 4}n_{\rm F}\right)
T^3$ \cite{kt}. Therefore, the ratios $\rho^{1 \over 4}/\bar E$ and
$n^{1 \over 3}/\bar E$ are measures for the deviation from thermal
equilibrium. For perturbative inflaton decay these ratios are
initially less than one, so that thermalization increases the number
density and reduces the mean energy. Complete thermalization therefore
requires interactions which change the number of particles to be in
equilibrium.

There are three types of interactions which help to build up and
maintain full (i.e. both kinetic and chemical) equilibrium: $2
\rightarrow 2$ scatterings, $2 \rightarrow N$ scatterings ($N \geq
3$), and particle decays. Elastic $2 \rightarrow 2$ scatterings
redistribute the energy between the scattered particles, but play no
role in achieving chemical equilibrium. Inelastic $2 \rightarrow 2$
(annihilation) reactions can help to maintain {\em relative} chemical
equilibrium between different particle species, but again leave the
total number of particles unchanged. For particles with energy $E \gg
T$ and number density $n(E)$, the rate of $2 \rightarrow 2$ reactions
with an energy exchange $\Delta E \sim E$ is typically given by
$\Gamma \sim \alpha^2 n(E) /E^2$, where $\alpha$ is the relevant
coupling constant. On the other hand, $2 \rightarrow N$ scatterings
and particle decays increase the number of particles and hence play a
crucial role in reaching full equilibrium. The rate of $2 \rightarrow
N$ reactions is suppressed by additional powers of $\alpha$. Therefore
inelastic scatterings with a large squared 4--momentum exchange $|t|
\sim E^2$ come to equilibrium later than $2 \rightarrow 2$ reactions
do. This would result in a late thermalization and a low reheat
temperature, if these were the most important reactions which increase
the total number of particles \cite{eeno,ee}. One possibility to
achieve chemical equilibrium more quickly is through the ``catalyzed
thermalization'' scenario \cite{mcdonald}. As shown in Refs.
\cite{mcdonald,a}, in the absence of interactions $n(E)/E^2$ increases
with decreasing $E$ in the spectrum of inflaton decay products, due to
redshifting of ``early'' inflaton decay products. This suggests that a
seed of particles with energy $E \ll m_\phi$, which constitute a tiny
fraction of the number density and the energy density of the plasma,
may thermalize much earlier than the bulk of the plasma. The large
number of created soft particles can then act as targets triggering a
rapid thermalization of the bulk. However, as pointed out in \cite{a},
elastic $2 \rightarrow 2$ scattering of particles in the bulk off
particles in the seed is efficient; it destroys the seed by bringing
it into kinetic equilibrium with the bulk. Nevertheless, catalyzed
thermalization can still take place if inflaton decay products
themselves decay before coming into kinetic equilibrium with the bulk
\cite{a}. However, it does not seem very probable that decays of
on-shell particles alone will be enough for building the chemical
equilibrium.

\setcounter{footnote}{0}

In general, inelastic scatterings which produce relatively soft
particles are the most important processes leading to full
equilibrium. This has recently been illustrated in Ref.\cite{ds} where
the $t-$channel scattering of two matter fermions with energy $\simeq
m_\phi/2$ (from inflaton decay) to two fermions, plus one gauge boson
with typical energy $E \ll m_\phi$, has been considered. The key
observation is that (for $T=0$ and in a flat space--time) $t-$channel
scattering is divergent as $\Delta E \rightarrow 0$. One thus has to
choose a physical infrared cut--off in order to estimate the
thermalization rate. A reasonable choice is the inverse of the average
separation between two particles in the plasma, $\bar r \propto
n^{-1/3}$, where $n$ is the number density of the plasma. This leads
to a thermalization rate for hard particles (with $E \sim m_\phi/2$)
of the order\footnote{We ignore factors of order $\log E/T \sim \log
m_\phi / T$ where $T$ is the temperature of the plasma, which could
increase the rate by a factor of a few. Note also that the authors of
Ref.\cite{ds} took $|t|_{\rm min} \sim T^2$, but estimated the target
density $n$ from the density of particles {\em before} thermalization,
$n \sim T^4 / m_\phi$. This results in a lower estimate of the
thermalization rate.}
\beq \label{eqrate}
\Gamma_{\rm in} \sim n \cdot \sigma(2 \rightarrow 3) \sim \alpha^3
n^{1/3}. 
\eeq
Once $\Gamma_{\rm in} > H$, relatively soft gauge boson with energy
$\gsim T$ are efficiently created. This increases $n$, and hence the
thermalization rate (\ref{eqrate}). Note that the increased target
density over--compensates the increase of the cut--off $|t|_{\rm
min}$. From then on the number of particles with energy $\sim T$
increases faster than exponentially and quickly reaches its final
value, as pointed out in Ref.\cite{es}\footnote{The situation is
slightly more subtle if the soft gauge boson belongs to an abelian
gauge group (e.g. photon). As we will show shortly, only inelastic
scatterings with gauge boson exchange in the $t-$ or $u-$channel
results in the estimate in eq.(\ref{eqrate}). Such diagrams exist for
scattering of a fermion (e.g. quark) from a soft non--abelian gauge
boson (e.g. gluon) \cite{es}. On the other hand, in inelastic
electron--photon scattering the $t-$channel diagram has an electron as
exchanged particle, and hence has a rate much smaller than that of
electron--electron scattering. However, soft photons with $E \sim T$
annihilate into soft $e^+e^-$ pairs at a rate which exceeds
(\ref{eqrate}) by a factor of $1/\alpha$. These soft $e^\pm$ can then
serve as targets for subsequent scatterings of hard electrons. The
thermalization rate, which is set by the rate of the slowest relevant
reaction, is then still approximately given by (\ref{eqrate}).}.
Inelastic $2 \rightarrow 2$ reactions then rapidly build up full
kinetic equilibrium. This suggests that $\Gamma_{\rm in}$ can in fact
be considered as thermalization rate, with $n$ being the original
(pre--thermalization) density, $n \sim g_* T^4 / (3 m_\phi)$, where
$g_*$ is the effective number of relativistic degrees of freedom; here
$T$ is the temperature (\ref{temp}) which the plasma would have if it
were thermalized.

This argument assumes that $2 \rightarrow 3$ scattering can
efficiently produce ``soft'' particles with energy $\gsim T$. The
production of even softer particles would not help in thermalizing the
plasma, since it would not slow down the parent particle appreciably,
and might even lead to a density of such very soft particles which
exceeds their thermal density. $2 \rightarrow 3$ reactions with all
virtual particles (in propagators) having virtualities of order
$n^{1/3}$ can indeed produce ``soft'' particles with energy $E_s \gg
n^{1/3}$, if these ``soft'' particles are nearly collinear with an
incident hard particle, the emission angle being of order
$\sqrt{n^{2/3}/ (m_\phi E_s)}$. One might wonder whether the emission of
such a very collinear particle is in fact physically distinguishable
from no emission at all. We believe this is the case. Note first of
all that the lower cut on the virtualities of all propagators implies
that the ``collinear'' particle is in fact not that close to the
emitted one in full phase space. Moreover, after a time of order
$1/\Gamma_{\rm in}$ there will be many such soft particles. Scattering
of an almost collinear ``soft$+$hard'' pair on a soft particle can
yield a final state with soft particles being emitted at a large angle
only if the ``soft'' particle in the initial state participates; the
``hard'' particle in the initial state can only scatter at a very
small angle. In other words, even before full thermalization, the
plasma allows to physically detect the presence of soft, collinear
particles in the ``beam'' of hard particles; the process that
``detects'' these particles also removes them from the ``beam''. This
happens at a time scale significantly shorter than $\Gamma_{\rm in}$,
since one only needs $2 \rightarrow 2$ reactions here.

In order to check whether the inflaton decay products can thermalize
before the completion of inflaton decay, we compute the maximum
temperature $T_{\rm max}$ of the thermal plasma; thermalization
occurs before inflaton decay completes iff $T_{\rm max} > T_{\rm R}$. 
We use eq.(\ref{temp}), with $H = \Gamma_{\rm in}$ from
(\ref{eqrate}), and
$\Gamma_{\rm d} = g_*^{1/4} T_{\rm R}^2 / M_{\rm Planck}$, resulting in
\beq \label{tmax}
T_{\rm max} \sim T_{\rm R} \left ( \alpha^3 \left( \frac {g_*}{3}
\right)^{1/3} { M_{\rm Planck} \over m_\phi^{1/3} T_{\rm R}^{2/3} }
\right )^{3/8},
\eeq 
which is somewhat higher\footnote{The two estimates coincide if
$\Gamma_{\rm d}$ saturates its upper bound of $m^3_\phi / M_{\rm
Planck}^2$.} than the estimate in Ref.\cite{ds}:
\beq \label{tmax2}
T_{\rm max} \sim T_{\rm R} \left (\alpha^3 g_*^{3/4} {M_{\rm Planck} \over
3 m_\phi}\right )^{1/2}.
\eeq
In any case it is reasonable to expect that the largest temperature of
the Universe after inflation is between the values estimated in
Eq.(\ref{tmax}) and Eq.(\ref{tmax2}). An interesting observation is
that $T_{\rm max}$, from Eq.(\ref{tmax}), grows $\propto T_{\rm
R}^{3/4}$. Even if $m_{\phi}$ is near its COBE--derived upper bound
\cite{cdo} of $\sim 10^{13}$ GeV, for a chaotic inflation model, and
$T_{\rm R}$ is around $10^9$ GeV (which saturates the gravitino bound)
$T_{\rm max}$ will exceed $T_{\rm R}$ if the coupling $\alpha^3 \gsim
10^{-8}$. This is easily accommodated for particles with gauge
interactions as $\alpha$ will be the gauge fine structure constant.
For inflationary models with smaller $m_\phi$, early thermalization
can be realized with even weaker couplings. Recall also that we assume
perturbative inflaton decays, which requires $T_{\rm max} <
m_\phi/2$. Together with eq.(\ref{tmax}), taking $\alpha \lsim 0.1$,
this implies $T_{\rm max} \leq 10^{11} \ (10^5)$ GeV for $T_{\rm R} =
10^9 \ (1)$ GeV; this bound is saturated for $m_\phi = 2 T_{\rm
max}$. This implies in particular that there will be no ``wimpzilla''
($m_\chi \sim 10^{12}$ GeV) \cite{ckr} production from {\em
thermalized} inflaton decay products, even if we use the higher
estimate (\ref{tmax}) for $T_{\rm max}$.

For $\Gamma_{\rm in} \geq H \geq \Gamma_{\rm d}$, fresh inflaton
decays will keep producing particles with energy $E \simeq m_\phi/2$,
but these particles quickly thermalize by scattering off a large
number of soft particles in the thermal plasma. The time scale for
this process is again given by eq.(\ref{eqrate}), where now $n^{1/3}
\sim T$. Inflaton decay effectively completes at $H \simeq
\Gamma_{\rm d}$ when the plasma temperature is $T_{\rm R}$. From then
on the Universe is radiation--dominated and its temperature will be
redshifted as $a^{-1}$.

\vspace*{8mm}
\begin{center}
\SetScale{0.6} \SetOffset(20,40)
\begin{picture}(350,100)(0,0)
\Line(75,90)(175,90) \Text(30,65)[l]{$f_1$}
\Vertex(175,90){3}
\Line(175,10)(75,10) \Text(30,0)[l]{$f_2$}
\DashLine(175,90)(175,10){5} 
\Line(175,90)(275,90) \Text(175,65)[r]{$f_3$}
\Vertex(175,10){3}
\Line(275,10)(175,10) \Text(175,0)[r]{$f_4$}
\end{picture}
\vspace*{-13mm}

{\large a) }
\end{center}

\vspace*{-8mm}

\begin{center}
\SetScale{0.6} \SetOffset(20,40)
\begin{picture}(350,200)(0,0)
\DashLine(75,170)(175,170){5} \Text(39,105)[r]{$\phi_1$}
\DashLine(175,170)(175,70){5}
\Vertex(175,170){3}
\DashLine(75,70)(175,70){5}  \Text(39,45)[r]{$\phi_2$}
\DashLine(175,170)(275,190){5} \Text(170,120)[l]{$\phi_3$}
\Vertex(175,70){3}
\DashLine(175,170)(275,150){5} \Text(170,90)[l]{$\phi_4$}
\DashLine(175,70)(275,90){5} \Text(170,60)[l]{$\phi_5$}
\DashLine(275,50)(175,70){5} \Text(170,30)[l]{$\phi_6$}
\end{picture}
\vspace*{-25mm}

{\large b)} \vspace*{5mm}
\end{center}

\noindent
{\bf Fig. 1:}~Typical scattering diagrams with scalar boson exchange in
the $t-$channel for: (a) fermions or antifermions in the initial and
final state, (b) inelastic $2 \rightarrow 4$ scattering involving only
scalar particles.
\vspace*{6mm}

Since the estimate for $T_{\rm max}$ crucially depends on the rate for
inelastic scatterings it is instructive to explore this issue a little
more deeply. As mentioned earlier, for processes with a $t$-channel
singularity, e.g. $e^+ e^- \rightarrow e^+ e^- \gamma$ via the
$t$-channel exchange of a photon, one finds $\sigma_{\rm in} \propto
|t|^{-1}$ rather than $\sigma_{\rm in} \propto s^{-1}$. In fact this
is the very same singularity which exists in $e^+ e^- \rightarrow e^+
e^-$ scattering in the $t-$channel. The appearance of $|t|^{-1}$ in
the cross section can be understood in the following way. $\sigma$
receives a factor of $|t|^{-2}$ from the photon propagator while the
contribution from the phase space integration is $\propto |t|$, thus
resulting in $\sigma \propto |t|^{-1}$. This implies that boson
exchange in the $t-$channel is required in order to have $\sigma
\propto |t|^{-1}$ since the contribution of diagrams with fermion
propagator will be $\propto s^{-1}$ (after phase space integration).
Moreover, it can be shown that for essentially massless external
particles the contribution from scalar boson exchange is also
suppressed. For example, consider the $t-$channel scattering of two
fermions via scalar exchange, as shown in Fig.~1a. A
fermion--fermion--scalar vertex naturally arises from a Yukawa
coupling or, in supersymmetric models, from the $D-$term part of the
action. Note that scalar interactions flip the chirality of the
fermion line. However, for massless fermions a flip of chirality also
implies a flip of helicity, i.e. a spin flip; this is forbidden by
angular momentum conservation for forward scattering, where $t
\rightarrow 0$. As a result, the diagram of Fig.~1a has no $t-$channel
singularity at all.\footnote{This can also be seen easily by computing
the relevant Dirac traces. Each fermion line gives rise to a separate
trace, which is proportional to the scalar product of the 4--momenta
of the two fermions involved. For massless particles these exactly
cancel the $1/t^2$ of the squared scalar propagator.}

Finally, it is also possible to have inelastic $2 \rightarrow 4$
scatterings, shown in Fig.~1b, with only scalar particles in external
and internal lines. A vertex comprising four scalars can for example
arise from the Higgs self--coupling in the standard model (SM), or
from the $F-$ and $D-$term parts of the scalar potential in
supersymmetric models. However, in this case integration over the
total allowed phase space only leads to logarithmic divergencies in
the limit of vanishing external masses. Including the squared
$t-$channel propagator in Fig.~1b, the integral in question can be
written as $\int dt d M^2_{34} d M^2_{56}\, 1/t^2$, where $M^2_{ij} =
(p_i+p_j)^2$ is the squared invariant mass of the pair $i,j = 3,4$ or
$5,6$. $|t|$ reaches its minimum for small $M^2_{ij}$, in which case
$|t|_{\rm min} \simeq (M^2_{34} M^2_{56})/s$; the integrals over
$M^2_{34}$ and $M^2_{56}$ then only give rise to logarithmic
singularities, as advertised. Ignoring such logarithmic factors, the
total cross section will thus again behave as $\sigma \propto s^{-1}$.
                          
The fact that only diagrams with light gauge bosons as internal and/or
external lines give rise to $\sigma_{\rm in} \propto |t|^{-1}$ can
have interesting implications for thermalization, in scenarios where
the gauge group is completely broken in the early Universe. This may
for example happen in supersymmetric models where flat directions in
the scalar potential \cite{gkm} can acquire a large vacuum expectation
value (VEV) during inflation\footnote{For a perturbative decay,
inflaton couplings to other fields are so small that the mass induced
by the inflaton VEV (if any) will be much smaller than the inflaton
mass. Therefore scalar fields can indeed acquire a large VEV during
inflation.} \cite{flat}. If gauge bosons have a mass $m_g$ such that
$\alpha^{-1} n^{1/3} < m_g < m_\phi$, then the most important diagrams
are those with resonant gauge boson exchange in the $s-$channel, with
$\sigma_{\rm in} \sim \alpha m^{-2}_{g}$. Moreover, for $m_g > m_\phi$
we will have $\sigma_{\rm in} \sim \alpha^3 m^{-2}_{\phi}$. This
affects the estimate for the thermalization rate $\Gamma_{\rm in}$ and
may indeed result in a (much) smaller $T_{\rm max}$. As one
consequence, thermal effects on the flat direction dynamics
\cite{ace,ad} may be alleviated. Once flat directions start
oscillating their VEV is redshifted, due to the expansion of the
Universe, and the induced mass for the gauge bosons rapidly
decrease. The flat direction oscillations starts when $H \sim V''$,
the second derivative of the scalar potential. In models where
supersymmetry breaking is communicated to the visible sector through
gauge interactions at a relatively low scale (of order tens of TeV)
the potential at high field values is exceedingly flat. It is
therefore conceivable that the flat direction induced masses for the
gauge bosons, if they completely break the gauge group, could delay
thermalization until late times. This might allow to construct models
with low reheat temperature even if the inflaton decay width
$\Gamma_{\rm d}$ is not very small. In the remainder of this article
we ignore such possible effects and use the (rather generous) estimate
in eq.(\ref{tmax}) for $T_{\rm max}$. This allows us to study the most
efficient production of massive particles which is possible from the
plasma of inflaton decay products.

\section{Heavy particle production}

\setcounter{footnote}{0} 

We now move on to the issue of heavy particle production. In recent
years several mechanisms have been put forward for creating very
heavy, even superheavy, particles in cosmologically interesting
abundances. For instance particle production could take place in a
time-varying gravitational background during the phase transition from
inflationary to the radiation-dominated or matter-dominated phase
\cite{ckr2}. Another possibility is to create supermassive particles
from preheating \cite{ckr3}. Here we will focus on production of very
massive particles from various processes, including a thermal bath,
during perturbative reheating. Note that particle production from
other sources, if present, would further strengthen the bounds which
we will derive as they simply add to production from mechanisms
discussed here. 

The situation for a weakly interacting massive particle (WIMP) $\chi$
with mass $m_\chi \lsim T_{\rm R}$ is very well established
\cite{jkg}. Such a species is at equilibrium with the thermal bath for
$T > m_\chi$ and its number density follows $n_\chi \propto T^3$. Once
the temperature drops below $m_\chi$, the particle becomes
non--relativistic and its equilibrium number density becomes $n^{\rm
eq}_{\chi} \sim {(T m_\chi)}^{3/2} {\rm exp}(-m_\chi/T)$. Pair
annihilation of $\chi$s to light particles, occuring at a rate
approximately given by $\Gamma_{\rm ann} \sim \alpha_\chi^2 n_\chi
m^{-2}_{\chi}$, preserves its chemical equilibrium with the bath so
long as $\Gamma_{\rm ann} \geq H$. Here $\alpha_\chi$ is the effective
coupling constant of $\chi$ to particles in the plasma. The
annihilation rate eventually drops below the Hubble expansion rate,
mostly due to the fact that $n_\chi^{\rm eq}$ is exponentially
suppressed. Then the comoving number density of $\chi$ will be frozen
at its final value $n_\chi \sim {(T_{\rm f} m_\chi)}^{3/2} {\rm
exp}(-m_\chi/T_{\rm f})$, where $T_{\rm f}$ denotes the freeze--out
temperature. In the post--inflationary era, i.e. for $H < \Gamma_{\rm
d}$, typically $T_{\rm f} \sim (m_\chi/20)$ if $\alpha_\chi \sim
\alpha$, up to logarithmic corrections \cite{jkg}. Taking $\alpha_\chi
\simeq \alpha (m_\chi/M_Z)^2$ gives a lower bound of a few GeV on the
mass of heavy, stable neutrinos, known as the Lee--Weinberg bound. In
the other extreme, the unitarity bound $\sigma(\chi\chi \rightarrow
{\rm anything}) < 4 \pi m_\chi^{-2}$ provides a firm upper bound of
about 100 TeV \cite{gk} on any stable particle that was in thermal
equilibrium at any temperature $T < T_{\rm R}$.\footnote{Provided the
comoving entropy of the Universe remained essentially constant for $T
< T_{\rm f}$.}

The upper bound on $m_\chi$ might be relaxed considerably if the
initial thermal equilibrium condition for $\chi$ is relaxed, i.e. if
$T_{\rm f} > T_{\rm R}$. Even in that case the $\chi$ particles can
have been in thermal equilibrium with the plasma of SM particles,
whose temperature $T$ can significantly exceed $T_{\rm R}$ at
sufficiently early times $t < 1/\Gamma_{\rm d}$. However, we saw above
that at those times the thermal bath did not dominate the energy
density of the Universe. This gives rise to a significant difference
from the freeze--out situation discussed above, due to the different
$T$ dependence of the redshift factor. In this case $\chi$ pair
annihilation reactions freeze out at a higher temperature, since $H
\sim \sqrt{g_*} T^4/(T^{2}_{\rm R} M_{\rm Planck}) > \sqrt{g_*}
T^2/M_{\rm Planck}$ for $H > \Gamma_{\rm d}$. Also, $n_\chi$ is now
redshifted $\propto T^8$, which is much faster than in a
radiation--dominated Universe. The situation in this case has been
investigated in detail in Refs.\cite{ckr,gkr,ky,rosenfeld} where the
relevant Boltzmann equations governing the production and annihilation
of $\chi$s are solved both numerically and analytically. In
Ref.\cite{ckr} out of equilibrium production of $\chi$ from
scatterings in the thermal bath is studied and the final result is
found to be (the superscript ``ss'' stands for $\chi$ production from
``soft--soft'' scattering; see below)
\bea  \label{ssrate1}
\Omega^{\rm ss}_\chi h^2 &\sim& \left ({200 \over g_*} \right )^{3/2}
\alpha_\chi^2 {\left ({2000 T_{\rm R} \over m_\chi} \right )}^7 
\nonumber \\ && \hspace*{2cm}
(\chi \ {\rm not \ in \ equilibrium}).
\eea
Here $\Omega_\chi$ is the $\chi$ mass density in units of the critical
density, and $h$ is the Hubble constant in units of 100
km$/$(s$\cdot$Mpc). We have taken the cross section for $\chi$ pair
production or annihilation to be $\sigma \simeq \alpha_\chi^2 /
m^{2}_{\chi}$. Most $\chi$ particles are produced at $T \simeq
m_\chi/4$ \cite{gkr}. The density of earlier produced particles is
strongly red--shifted, while $\chi$ production at later times is
suppressed by the Boltzmann factor. It is important to note that
$\Omega_\chi$ is only suppressed by $(T_{\rm R}/m_\chi)^7$ rather than
by $\exp(-m_\chi/T_{\rm R})$. We thus see that a stable particle with
mass $m_\chi \gg 100$ TeV might act as the Dark Matter in the Universe
(i.e. $\Omega_\chi \simeq 0.3$), if $m_\chi \sim 2000~T_{\rm R}
\cdot \alpha_\chi^{2/7}$. Recall that for $m_\chi < 20 T_{\rm R}$ the
standard analysis holds as the freeze--out occurs in a
radiation--dominated Universe.

Eq.(\ref{ssrate1}) predicts a relic density that increases with the
$\chi$ coupling strength. However, this is true only if the $\chi$
density is always smaller than its equilibrium density, which requires
\cite{gkr}
\beq \label{equilcon}
\alpha_\chi^2 \leq \bar \alpha_\chi^2 \simeq 200 \left( \frac
{g_*} {200} \right)^{1/2} \frac {m^3_\chi} {M_{\rm Planck} T^2_{\rm
R}}.
\eeq
If this condition is violated, today's $\chi$ relic density is given
by
\bea \label{ssrate2}
\Omega^{\rm ss}_\chi h^2 &\sim& \left( \frac {200} {g_*} \right)^{1/2}
\frac {T_{\rm R} x_{\rm f}^{4+a}} { m_\chi \alpha_\chi^2}  \left( \frac
{T_{\rm R}} { 8 \cdot 10^5 \ {\rm GeV}} \right)^2 \nonumber \\ &&
\hspace*{2cm} (\chi \ {\rm in \ equilibrium}) ,
\eea
where the exponent $a = 0 \ (1)$ if $\chi \chi$ annihilation proceeds
from an $S$ ($P$) wave initial state. The freeze--out temperature is
now given by $x_{\rm f} \equiv m_\chi/ T_{\rm f} \simeq \log ( 0.08
g_*^{-1/2} \alpha_\chi^2 x_{\rm f}^{2.5-a} M_{\rm Planck} T^2_{\rm R}
/ m^3_\chi)$, as compared to $x_{\rm f} \sim \log ( 0.2 g_*^{-1/2}
\alpha_\chi^2 x_{\rm f}^{0.5-a} M_{\rm Planck} / m_{\chi})$ if
freeze--out occurs at $T < T_{\rm R}$.

Of course, eqs.(\ref{ssrate1}) and (\ref{ssrate2}) are only applicable
if $m_\chi \lsim 2 T_{\rm max}$, cf. eq.(\ref{tmax}) and the subsequent
discussion. This requires
\bea \label{sscon}
\frac {m_\chi} {T_{\rm R}} &<& 4 \alpha^{9/8} \left( \frac {g_*} {200}
\right)^{1/8} \frac { M_{\rm Planck}^{3/8} } { T_{\rm R}^{1/4}
m_\phi^{1/8} } 
\nonumber \\
&\leq& 4 \alpha \left( \frac {g_*} {200} \right)^{1/9}
\left( \frac {M_{\rm Planck}} {T_{\rm R}} \right)^{1/3};
\eea
in the second step we have used the condition $T_{\rm max} \leq m_\phi
/ 2$. For example, for $\alpha = 0.05$, $T_{\rm max} \gsim 1000 T_{\rm
R}$ is only possible if $T_{\rm R} < 2 \cdot 10^{-12} M_{\rm
Planck}$. Eq.(\ref{equilcon}) shows that equilibrium is more difficult
to achieve for larger $m_\chi$; this is not surprising, since the
cross section for $\chi$ pair production scales like $m_\chi^{-2}$.
This means that eq.(\ref{ssrate1}) will usually be applicable in
models with large $m_\chi$ and comparatively smaller $T_{\rm R}$,
while in the opposite case eq.(\ref{ssrate2}) may have to be used. We
will see shortly that both situations can arise in potentially
interesting scenarios. Finally, for fixed $T_{\rm R}$ and $m_\chi$,
$\chi$ production from the thermal plasma is maximized if the
condition (\ref{equilcon}) is saturated. This gives
\beq \label{maxssrate}
\Omega^{\rm ss,max}_\chi h^2 \sim 3 \cdot 10^{25} \frac {200} {g_*}
 \frac {T_{\rm R}^5} {m_\chi^4 M_{\rm Planck}}.
\eeq

So far we have only discussed $\chi$ production from the scattering of
``soft'' particles in the thermal distribution. However, ``hard''
particles with energy $E \simeq m_\phi/2 \gg T$ are continuously
created by (two body) inflaton decay for $H \geq \Gamma_{\rm
d}$. These particles eventually thermalize with the bath, but this
takes a finite amount of time. The presence of hard inflaton decay
products can therefore affect heavy particle production in two
ways. Firstly, $\chi$s can be produced from $2 \rightarrow 2$
scatterings of a hard particle off either soft particles in the
thermal bath (if kinematically allowed), or off other hard particles.
Moreover, $\chi$s might be directly produced from inflaton decay. In
both cases $\chi$ production may be enhanced and this is the issue to
which we now turn.

\subsection{Particle production from hard--soft scatterings}

It is seen from eq.(\ref{infden}) that the energy density in inflaton
decay products created in the time interval $[(3/2)H^{-1},3H^{-1}]$,
for $H > \Gamma_{\rm d}$, is comparable to the energy density in the
existing thermal plasma which has a temperature $T \sim {(H
\Gamma_{\rm d} M_{\rm Planck})}^{1/4}$. This implies that
thermalization of these hard particles will increase the comoving
number density of the thermal bath, roughly speaking by a factor of 2,
in addition to bringing the decay products into kinetic equilibrium.
To the accuracy we are working in, we can therefore set the density of
``hard'' inflaton decay products produced during that time
interval\footnote{More exactly, $\bar n_h = \int_t^{2t} d n_h / dt$,
where only the {\em production} of hard particles from inflaton decay
is included in $d n_h / d t$, i.e. $d n_h / dt = \Gamma_{\rm d}
n_\phi(t)$.} to $\bar n_h \sim \rho/m_\phi \sim 0.3 g_* T^4/m_\phi$.
In order to estimate the rate of heavy particle production from these
``hard'' inflaton decay products, we also have to know the time needed
to reduce the energy of these hard particles from a value $\sim
m_\phi/2$ to a value near $T$. As shown in ref.\cite{ds}, $2
\rightarrow 2$ scattering reactions are {\em not} very efficient. The
reaction rate is large, but the average energy loss per scattering is
only ${\cal O}(T^2/m_\phi)$, giving a slow--down time of order $
\left[ \alpha^2 T^2 / m_\phi \right]^{-1}$ (up to logarithmic
factors). On the other hand, we saw at the end of Section 2 that
inelastic $2 \rightarrow 3$ reactions allow large energy losses (in
nearly collinear particles) even if all virtual particles only have
virtuality of order $T$. The slow--down rate is thus again given by
eq.(\ref{eqrate}), where the ``target'' density is now $n \simeq 0.2
g_* T^3$:\footnote{We note in passing that the logarithmic factors
cancel in eq.(\ref{slowrate}). We included the $g_*$ factor in the
cut--off $|t|_{\rm min}$ since (almost) all particles in the plasma
have non--abelian interactions with the exchanged particles in the $2
\rightarrow 3$ scattering diagram.}
\beq \label{slowrate}
\Gamma_{\rm slow} \simeq 3 \alpha^3 T \left( \frac {g_*} {200}
\right)^{1/3}. 
\eeq

Next let us estimate the rate for $\chi$ pair production from
hard--soft scatterings. This process is kinematically allowed so long
as $ET \geq 4 m^{2}_{\chi}$, where $E$ is the energy of the hard
particle so that the square of the center--of--mass energy is
typically a few times $ET$. The hard particle initially has energy $E
\simeq m_\phi/2$ and average (see above) number density $\bar n_h \sim
g_* T^4/(3 m_\phi)$, just after its production from inflaton decay. It
looses its energy at the rate given in eq.(\ref{slowrate}). Note that
the time required for the hard particle to drop below the kinematical
threshold $E_{\rm min} \simeq m^2_\chi / T$ depends only
logarithmically on this threshold. We ignore this logarithmic factor,
and simply estimate the slow--down time as $1/\Gamma_{\rm slow}$. On
the other hand, the rate for $\chi$ production from hard--soft
scatterings is approximately given by
\beq \label{hsprod}
\Gamma^{\rm hs}_\chi \sim \left( \frac {\alpha^2_\chi} {T m_\phi}
+ \frac {\alpha \alpha_\chi^2} {m_\chi^2} \right) 0.2 T^3,
\eeq 
where we conservatively assume that $\chi$'s can be produced from
scatterings of just one species (e.g. electrons). The two
contributions in (\ref{hsprod}) describe $2 \rightarrow 2$ reactions
with squared center--of--mass energy $\sim m_\phi T$ and ``radiative
return'' $2 \rightarrow 3$ reactions, respectively; in the latter
case the hard particle emits a collinear particle prior to the
collision, thereby reducing the effective cms energy of the collision
to a value near $m_\chi$. This results in the estimate\footnote{More
precisely, this is the density of $\chi$ particles produced between
$t$ and $2t$, i.e. during one Hubble time, during which $T$ remained
approximately constant.}
\beq \label{hsrate1}
n_\chi^{\rm hs}(T) \sim \bar n_h \cdot \frac {\Gamma^{\rm hs}_\chi}
{\Gamma_{\rm slow}} \sim 4 \left( \frac {g_*} {200} \right)^{2/3} \frac
{\alpha_\chi^2} {\alpha^2} \left( \frac{T^5}{\alpha m^2_\phi} + \frac
{T^6} {m_\phi m_\chi^2} \right), 
\eeq
for $\chi$s produced at temperature $T$. It is clear that because of
the redshift factor $(T_{\rm R}/T)^8$ production close to $T_{\rm
thr}$ makes the dominant contribution, where $T_{\rm thr} \equiv 4
m^{2}_{\chi}/m_\phi$. In order to make a safe (under)estimate we
choose the temperature $T_0 = 2 T_{\rm thr}$ for presenting our
results; note that the $\chi$ pair production cross section at
threshold, $s = 4 m^2_\chi$, is suppressed kinematically. If $T_0 <
T_{\rm R}$, the physical $\chi$ density at $H = \Gamma_{\rm d}$,
i.e. at $T = T_{\rm R}$, is simply given by eq.(\ref{hsrate1}) with $T
= T_{\rm R}$. In this case the maximal cms energy at $T_{\rm R}$ is
still above $8 m^2_\chi$. This means that the total $\chi$ production
cross section will be dominated by $2 \rightarrow 3$ ``radiative
return'' reactions, i.e. the second, $T^6$ contribution in
eq.(\ref{hsrate1}) will usually dominate. On the other hand, $T_0 >
T_{\rm R}$ leads to a physical $\chi$ density at $H = \Gamma_{\rm d}$
of order
\beq \label{hsrate2}
n_\chi^{\rm hs}(T_R) \sim 10^{-2} \left( \frac {g_*} {200}
\right)^{2/3} \frac {\alpha_\chi^2} {\alpha^3} \frac {T_{\rm R}^8
m_\phi} {m_\chi^6}, 
\eeq
In this case the contribution from $2 \rightarrow 3$ processes is
suppressed (by an extra power of $\alpha$).

In order to translate the $\chi$ density at $T = T_{\rm R}$ into the
present $\chi$ relic density, we use the relation \cite{kt}
\bea \label{relden}
\Omega_\chi h^2 &=& \frac { m_\chi n_\chi(T_{\rm R}) } { \rho_R (T_{\rm
R})} \cdot \frac {T_{\rm R}} {T_{\rm now}} \cdot (\Omega_R h^2)_{\rm
now}
\nonumber \\
&=& 6.5 \cdot 10^{-7}\, \cdot \, \frac {200} {g_*} \cdot \frac
{m_\chi n_\chi(T_{\rm R}) } {T_{\rm R}^3 T_{\rm now}},
\eea
where $\rho_R$ is the energy density in radiation, and we have used
$(\Omega_R h^2)_{\rm now} = 4.3 \cdot 10^{-5}$ \cite{ckr}. Our final
results for the contribution of hard--soft collisions to the $\chi$
relic density are thus: for $T_0 < T_{\rm R}$:
\bea \label{hsrate3}
\Omega_\chi^{\rm hs} h^2 \sim && \left( \frac {200}{g_*}
\right)^{1/3} \frac {\alpha_\chi^2}{\alpha^2} \left( \frac {T_{\rm R}}
{10^4 \ {\rm GeV} } \right)^2 \frac {10^{13} \ {\rm GeV}} {m_\phi}
\nonumber \\ &\cdot&
\frac {100 T_{\rm R}} {m_\chi} \left( 1 + \frac {m^2_\chi} {\alpha
T_{\rm R} m_\phi} \right), \ \ \ (T_0 < T_{\rm R})
\eea
where the second term in the last round parentheses comes from $2
\rightarrow 2$ processes. In the opposite situation, we have
\bea \label{hsrate4}
\Omega_\chi^{\rm hs} h^2 &\sim& \left( \frac {200}{g_*} \right)^{1/3}
\frac {\alpha^2_\chi} {\alpha^3} \frac {m_\phi} {10^{13} \ {\rm GeV}}
\left( \frac {3000 T_R} {m_\chi} \right)^5, \nonumber \\
&& \hspace*{2cm} (T_0 > T_{\rm R}) 
\eea
where we have again ignored the contribution from $2 \rightarrow 3$
processes. 

Two conditions have to be satisfied for our estimates
(\ref{hsrate1})--(\ref{hsrate4}) to be applicable. First, we need $T_0
< T_{\rm max}$. Note that this constraint is weaker by a factor
$m_\chi / m_\phi$ than the analogous constraint for the applicability
of eqs.(\ref{ssrate1}), (\ref{ssrate2}). Second, at temperature $T =
\max(T_{\rm R}, T_0)$, $n_\chi$ should be sufficiently small that
$\chi \chi$ annihilation reactions can be ignored. This is true if the
$\chi\chi$ annihilation rate is smaller than the expansion rate,
$\alpha_\chi^2 n_\chi / m_\chi^2 < H(T) \simeq \sqrt{g_*} T^4 /
(M_{\rm Planck} T_{\rm R}^2)$. If $T_0 < T_{\rm R}$, $\chi \chi$
annihilation is thus negligible if
\beq \label{hscon1}
\left( \frac {m_\chi} {T_{\rm R}} \right)^4 > \frac{1}{3} \left( \frac
{g_*} {200} \right)^{1/6} \left( \frac {\alpha_\chi^2} {\alpha}
\right)^2 \frac {M_{\rm Planck}} {m_\phi}. \ \ \ (T_0 < T_{\rm R})
\eeq
Recall that we need $m_\chi > 20 T_{\rm R}$, since otherwise $\chi$'s
would have been in equilibrium at $T_{\rm R}$. Moreover, typically
$\alpha_\chi^2 / \alpha \lsim 10^{-2}$ for weakly interacting
particles. The condition (\ref{hscon1}) will therefore always be
comfortably satisfied in chaotic inflation, where $m_\phi \sim 10^{-5}
M_{\rm Planck}$. If $T_0 > T_{\rm R}$, the condition for $\chi\chi$
annihilation to be negligible is
\beq \label{hscon2}
\frac{1}{3} \left( \frac {g_*} {200} \right)^{1/6} \frac
{\alpha^4_\chi} {\alpha^3} < \frac {m_\phi^3} {T_{\rm R}^2 M_{\rm
Planck} }. \ \ \ (T_0 > T_{\rm R})
\eeq
Again, this condition can only be violated if $m_\phi \ll 10^{13}$
GeV. We thus conclude that in most scenarios with rather heavy
inflatons the estimates (\ref{hsrate1})--(\ref{hsrate4}) are indeed
applicable. 

Finally, unless $m_\chi$ is quite close to $m_\phi$, in which case
$T_0$ is likely to exceed $T_{\rm max}$, $T_0$ is well below $m_\chi$;
if $m_\chi > 20 T_{\rm R}$, $T_0$ is then also well below the
freeze--out temperature $T_{\rm f}$ even if $\chi$'s once were in
thermal equilibrium with the plasma. The $\chi$ density from
hard--soft scattering can thus simply be added to the contribution
from soft--soft scattering.

For a first assessment of the importance of the contribution from
hard--soft scattering, we compare eq.(\ref{hsrate3}) with the maximal
contribution (\ref{maxssrate}) from soft--soft scattering. We find
that the contribution from hard--soft scattering would dominate if
$m_\phi/m_\chi < 5 \cdot 10^{-5} \cdot (\alpha_\chi / \alpha)^2 \cdot
(m_\chi / T_{\rm R})^2$, where we have taken $g_* \simeq 200$ (as in
the MSSM). This condition can only be satisfied if $m_\chi \gg T_{\rm
R}$ is not too far below $m_\phi$, in which case eq.(\ref{maxssrate})
is not applicable anyway, since $T_{\rm max} < m_\chi$. In other
words, whenever the soft--soft contribution is near its maximum, it
will dominate over the hard--soft contribution. On the other hand, the
hard--soft contribution will often dominate over the soft--soft one if
$\chi$ never was in thermal equilibrium. For $T_0 < T_{\rm R}$, this
is true if $\alpha^2 m_\phi / m_\chi < (m_\chi / T_{\rm R})^4 (m_\chi /
10^{16} \ {\rm GeV})$, which is satisfied unless $T_{\rm R}$ and
$m_\chi$ are both rather small. Similarly, for $T_0 > T_{\rm R}$,
eq.(\ref{hsrate4}) will dominate over eq.(\ref{ssrate1}) if $m_\phi >
10^{13} \ {\rm GeV} \cdot \alpha^3 (700 T_{\rm R} / m_\chi)^2$. Recall
that the last term must be $\lsim 0.1$, since otherwise the soft--soft
contribution by itself would overclose the Universe. In this case
the hard--soft contribution will thus dominate in models with rather
large inflaton mass. Of course, this means that cosmological
constraints on the model parameters $T_{\rm R}$ and $m_\chi$ will
often be considerably stronger than previously thought. We will come
back to this point when we present some numerical examples.

\subsection{Particle production from hard--hard scatterings}
\setcounter{footnote}{0}

If $T_0 > T_{\rm R}$, we should also consider $\chi$ production from
scattering of two hard particles. Recall that ``hard'' particles are
continuously created as long as $H > \Gamma_{\rm d}$. Collisions of
these particles with each other can produce $\chi$ pairs if $m_\chi <
m_\phi / 2$. Note that this constraint is independent of the
temperature. On the other hand, once a thermal plasma has been
established, the density of ``hard'' particles will always be much
smaller than the density of particles in the plasma. If hard--soft
scattering at $T = T_{\rm R}$ can still produce $\chi$ pairs, it will
certainly dominate over hard--hard production of $\chi$'s. On the
other hand, it's also possible that $T_0 > T_{\rm max}$, in which case
hard--soft scattering (and soft--soft scattering) does not produce any
$\chi$ particles.

Note that the rate of $\chi$ production from hard--hard scattering is
quadratic in the density of hard particles. We can no longer use our
earlier approximate solution of the Boltzmann equation in terms of the
density $\bar n_h$ of hard particles produced in the time interval
from $t$ to $2t$, since the actual density $n_h(t)$ at any given time
will be much smaller than this. Let us first consider the era after
thermalization, where a plasma with temperature $T$ already exists. As
noted earlier, a hard particle will then only survive for a time $\sim
1/\Gamma_{\rm slow}$, see eq.(\ref{slowrate}). During that era,
i.e. for $T_{\rm max} > T > T_{\rm R}$, the production of hard
particles from inflaton decays and their slow--down will be in
equilibrium, i.e. the instantaneous density $n_h(t) = 2 \Gamma_d
n_\phi(t) / \Gamma_{\rm slow}$, where $n_\phi$ is the density of
inflatons. This latter quantity is given by $2 \Gamma_{\rm d} n_\phi
\sim H g_* T^4 / (3 m_\phi)$, where we have made use of the fact that
$T$ doesn't change too much over one Hubble time. This leads to a
physical $\chi$ density at $T= T_{\rm R}$, for $\chi$ particles
produced during one Hubble time after thermalization, of order:
\beq \label{hh1}
n_\chi^{\rm hh}(T_R) \sim \frac {2000 T^2 T_{\rm R}^6
\sigma_{\chi\chi}} {\alpha^6 M_{\rm Planck} m_\phi^2} \left( \frac
{g_*} {200} \right)^{11/6}.
\eeq
We make the following ansatz for the $\chi$ production cross section:
\beq \label{sigma}
\sigma_{\chi \chi} \sim \frac {\alpha_\chi^2} {m_\phi^2} + \min \left[
\frac {\alpha \alpha_\chi^2} {m_\chi^2}, \ \alpha^2_\chi \alpha n_{\rm
plasma}^{-2/3} \right] . 
\eeq
The first term is again the perturbative $2 \rightarrow 2$ production
cross sections, whereas the second term cuts off the $t-$channel
propagator at the appropriate power of the density of relativistic
particles. This second term is needed since eq.(\ref{hh1}) shows that
$\chi$ production from hard--hard scattering is dominated by {\em
early} times, i.e. {\em high} temperatures. In particular, $T >
m_\chi$ is possible, in which case no propagator should be allowed to
be as large as $1/m_\chi^2$.

The fact that the highest temperatures give the biggest contribution
in eq.(\ref{hh1}) raises the question what happened before
thermalization. In this epoch the hard particles by definition didn't
have time to slow down appreciably (except by
red--shifting). Moreover, the co--moving inflaton density remained
essentially constant during that period. Hence now $n_h(t) \simeq 2
\Gamma_{\rm d} (t - t_{\rm I}) n_\phi(t)$, with $n_\phi(t) =
n_\phi(t_{\rm I}) (t_{\rm I}/t)^2 \simeq n_\phi(T_{\rm R}) /(t
\Gamma_{\rm d})^2 \simeq 1/t^2 \cdot M^2_{\rm Planck} / (6
m_\phi)$. In the last step we have used $n_\phi(T_{\rm R}) \simeq g_*
T_{\rm R}^4 / (3 m_\phi)$ and $\Gamma_{\rm d} \simeq \sqrt{g_*}
T^2_{\rm R} / M_{\rm Planck}$. Introducing $X_\chi = t^2 n_\chi$ (so
that $X_\chi$ is not affected by the Hubble expansion), the production
of $\chi$ particles before thermalization is described by
\beq \label{xeq}
\frac {d X_{\chi}} {d t} \simeq \sigma_{\chi \chi} \frac { g_*
M^2_{\rm Planck} T^4_{\rm R} } {m^2_\phi} \cdot \frac { (t - t_{\rm
I})^2 } {t^2},
\eeq
where the factor $T^4_{\rm R}$ comes from the factor $\Gamma_{\rm
d}^2$ contained in $n^2_h(t)$. For $t \gg t_{\rm I}$, $X_\chi$ will
thus grow linearly with $t$, which means that the physical density
$n_\chi \propto 1/t$. Of course, this behavior persists only until
the onset of thermalization, i.e. for $t \leq 1/\Gamma_{\rm in}$, see
eq.(\ref{eqrate}). Using eq.(\ref{tmax}) together with the requirement
$T_{\rm max} \leq m_\phi / 2$, it is easy to see that the solution to
eq.(\ref{xeq}) at $t = 1/ \Gamma_{\rm in}$ always exceeds the result
(\ref{hh1}). In other words, $\chi$ production through hard--hard
scattering is most efficient {\em before} thermalization. This is
perhaps not so surprising, since during this early epoch, the hard
particles have a higher physical density (once $t \gg t_{\rm I}$) and
survive longer than after thermalization has occurred. Including the
redshift from $T_{\rm max}$ to $T_{\rm R}$ and using eq.(\ref{relden})
we arrive at our final estimate
\beq \label{hhrate}
\Omega_\chi^{\rm hh} h^2 \sim 6 \cdot 10^{27} \cdot \left( \frac
{g_*}{200} \right)^{1/2} \sigma_{\chi \chi} \frac {m_\chi T_{\rm R}^7}
{m^2_\phi T^4_{\rm max}},
\eeq
where $\sigma_{\chi \chi}$ is given by eq.(\ref{sigma}) with $n_{\rm
plasma} \sim g_* T_{\rm max}^4 / (3 m_\phi)$. We have checked that for
$m_\phi$ near $10^{13}$ GeV the $\chi$ density from hard--hard
scattering always stays sufficiently small for $\chi$ annihilation
reactions to be negligible. It is also easy to see that for $T_0 <
T_{\rm R}$ the contribution (\ref{hhrate}) is less than the hard--soft
contribution (\ref{hsrate3}). However, the hard--hard contribution
will exceed the hard--soft contribution (\ref{hsrate4}) if $T_0 >
T_{\rm R}$ and $m_\chi \gsim 0.1 (m_\phi^5 T_{\rm R}^2 M_{\rm
Planck})^{1/8}$; this could indeed be the case if $m_\chi$ is rather
close to $m_\phi$ but well above $T_{\rm R}$.

\subsection{Numerical examples}

As well known, any stable particle must satisfy $\Omega_\chi < 1$,
since otherwise it would ``overclose'' the Universe. However, in some
cases other considerations give stronger constraints on the abundance
of $\chi$. This can happen, for example, for unstable massive
particles whose lifetime $\tau_\chi$ is comparable to the age of the
Universe $\tau_{\rm U}$. Radiative decays of such relics which take
place after recombination (i.e. at $t \geq 10^{13}$ s) are tightly
constrained by astrophysical bounds on the gamma--ray background
\cite{gamma}. The tightest bound arises for decays around the present
epoch \cite{gamma}:
\beq \label{decbound}
\Omega_\chi h^2 \leq 10^{-8}.
\eeq
Another interesting example is that of charged stable particles whose
abundance is severely constrained from searches for exotic isotopes
in sea water \cite{ky}. The most stringent bound on the abundance of
such particles with electric charge $-1$ is derived for masses $m_\chi
\simeq 100-~10000$ GeV \cite{ky}:
\beq \label{champsbound}
\Omega_\chi h^2  \leq 10^{-20};
\eeq
for heavier particles this bound becomes weaker. 

Having mentioned the different cosmological and astrophysical
constraints on long-lived or stable massive particles, in Fig.~2 we
present three numerical examples to compare the significance of
hard--soft and hard--hard scatterings with that of soft--soft
scatterings. We plot $\Omega_\chi h^2$ as a function of $m_\chi$ for
$\alpha_\chi = 0.01$ and $\alpha = 0.05$. The parameters $(T_{\rm R},
m_\phi)$ are chosen as $(10^8 \ {\rm GeV}, 10^{13} \ {\rm GeV})$ (a),
$(10^5 \ {\rm GeV}, 10^{13} \ {\rm GeV})$ (b) and $(3 \ {\rm MeV},
10^8 \ {\rm GeV})$ (c), respectively.

\setcounter{figure}{1}
\begin{figure}[htbp]
\vspace*{-.5cm}
\epsfig{figure=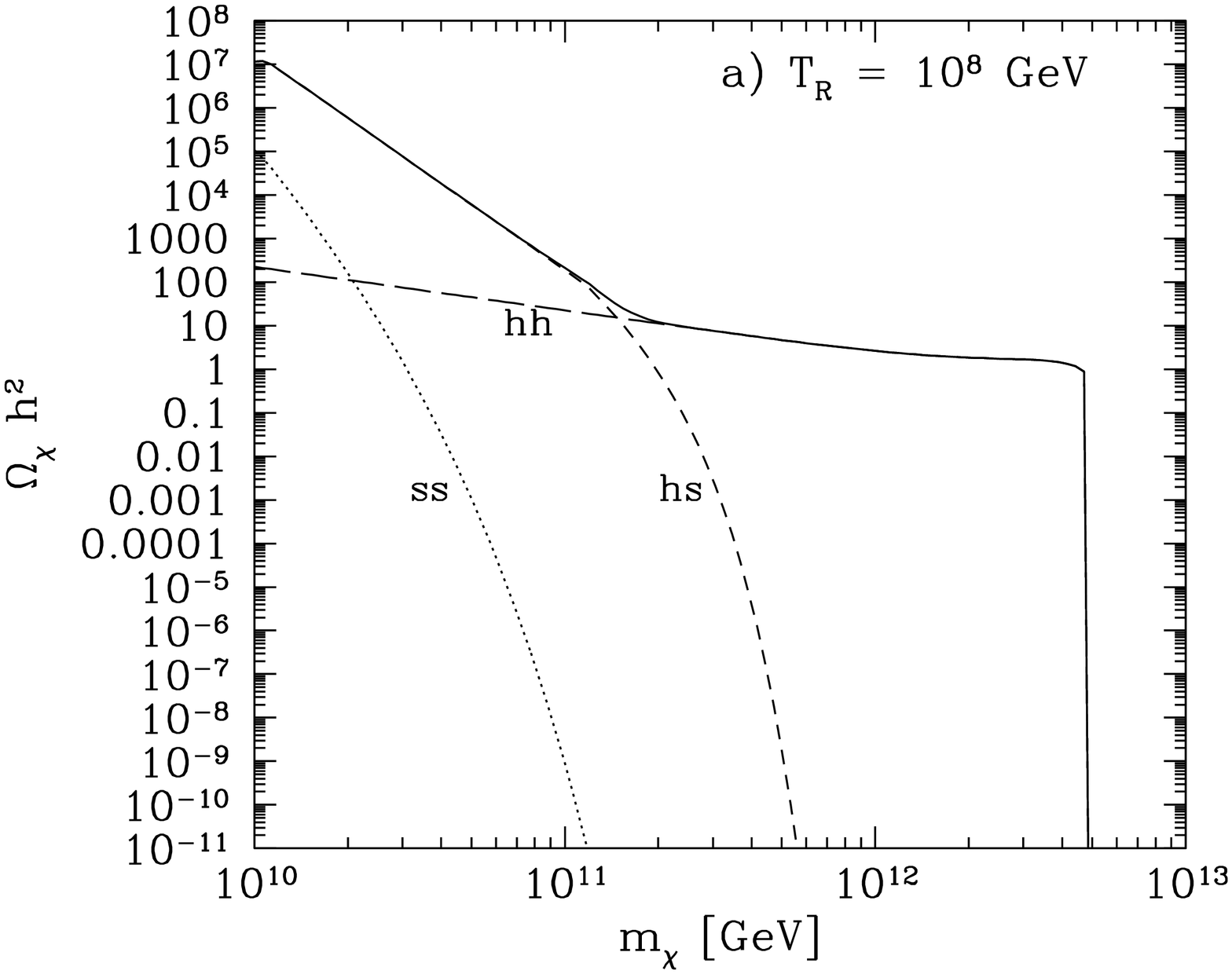,angle=-90,width=.54\textwidth,clip=}

\vspace*{-.5cm} 
\noindent
\epsfig{figure=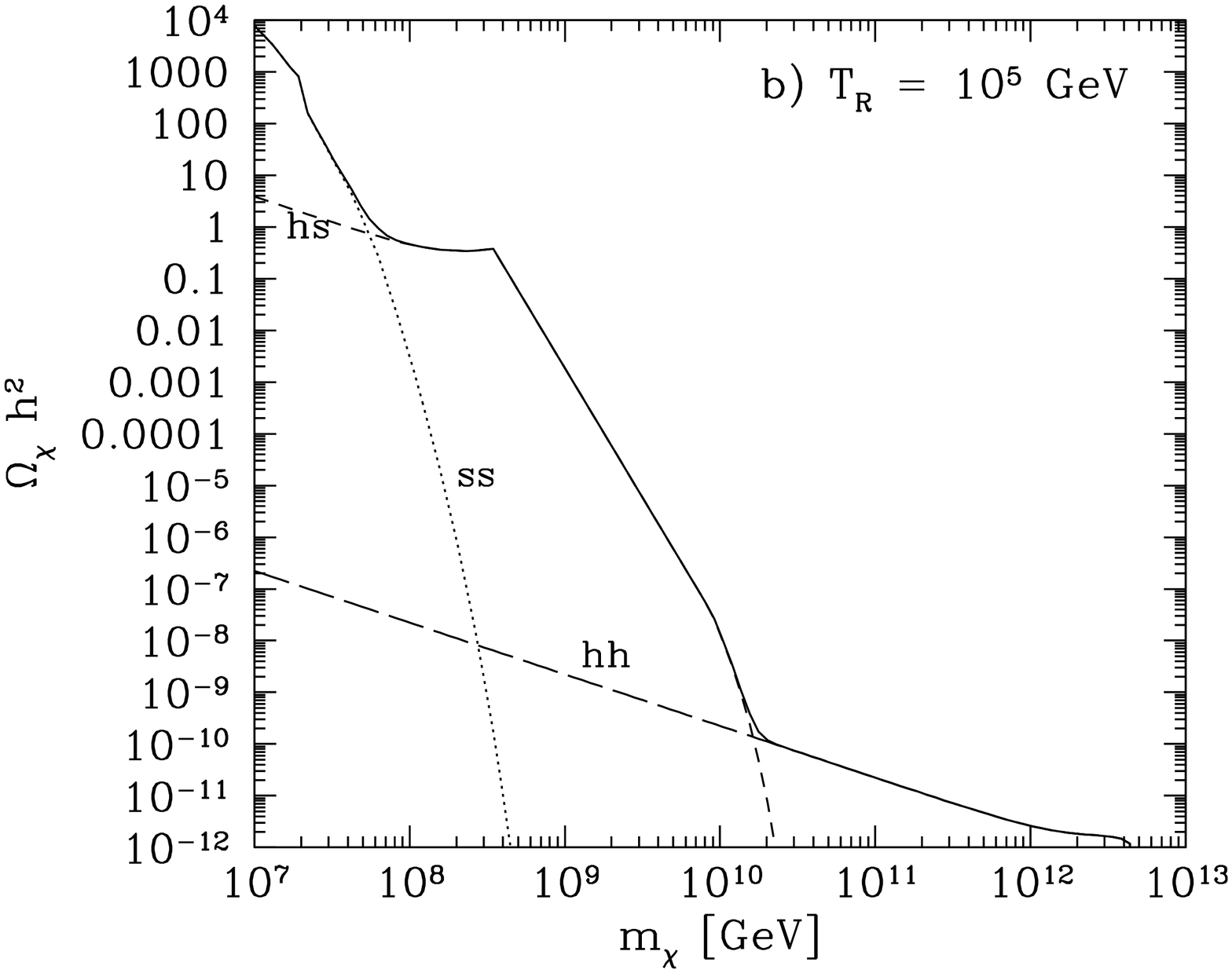,angle=-90,width=0.54\textwidth,clip=}

\vspace*{-.5cm} 
\noindent
\epsfig{figure=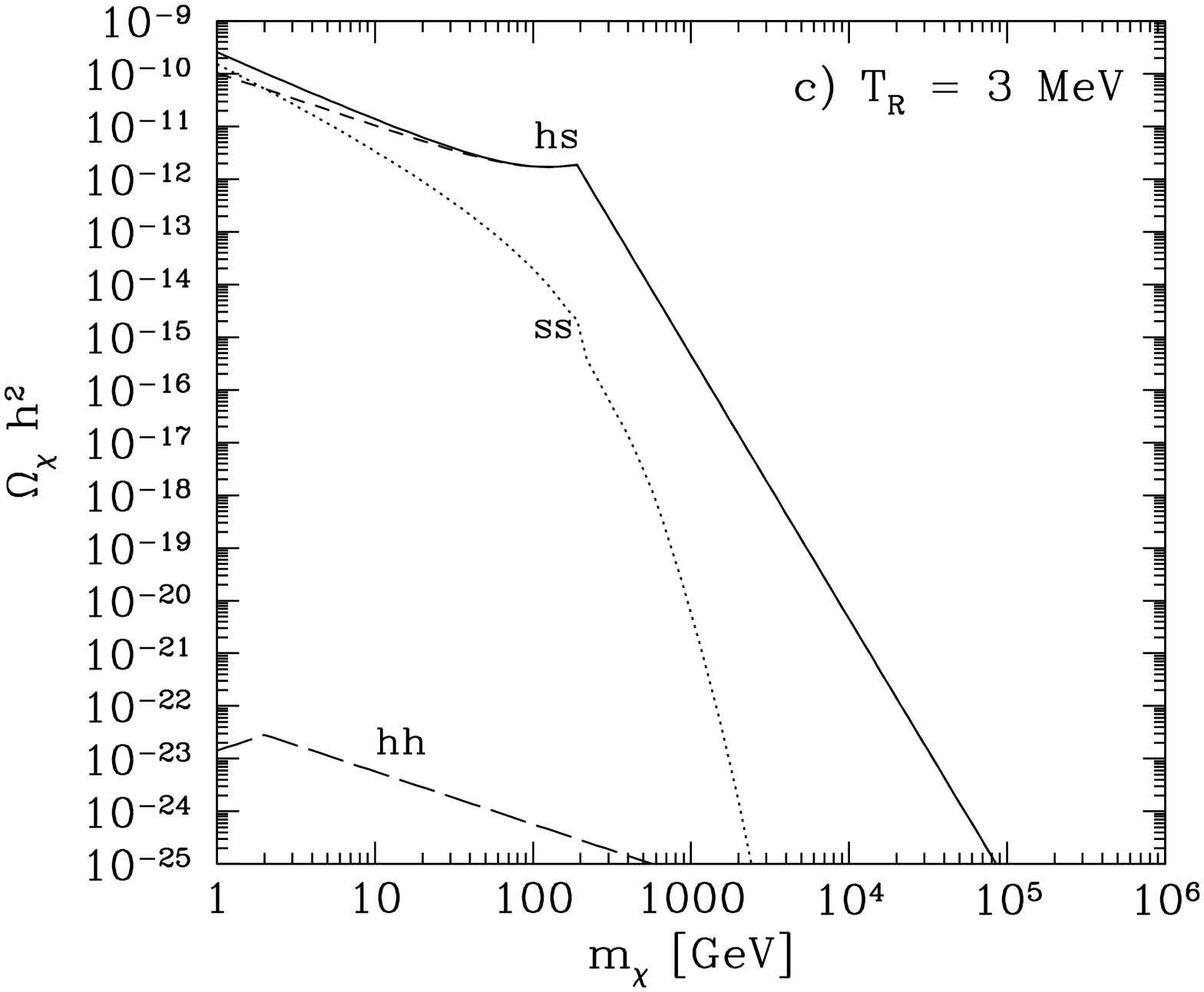,angle=-90,width=0.54\textwidth,clip=}

\caption{The relic density $\chi$ particles would currently have if
they are absolutely stable is shown as a function of $m_\chi$
for couplings $\alpha = 0.05, \ \alpha_\chi = 0.01$, and a) $(T_{\rm R},
m_\phi) = (10^8 \ {\rm GeV}, 10^{13} \ {\rm GeV})$, b)$(10^5 \ {\rm
GeV}, 10^{13} \ {\rm GeV})$, c)$ (3 \ {\rm MeV}, 10^{8} \ {\rm
GeV})$. The soft--soft, hard--soft and hard--hard contributions are
shown by the dotted, short dashed and long dashed curves,
respectively, while the solid curves show the sum of all three
contributions.} 
\end{figure}

In Fig.~2a, $T_{\rm max} \simeq 2 \cdot 10^{10}$ GeV; this explains
the very rapid fall of soft--soft contributions (which are never in
equilibrium for the range of $m_\chi$ shown) for $m_\chi > 3 \cdot
10^{10}$ GeV. We have multiplied the result of eq.(\ref{ssrate1}) with
$\exp(-2 m_\chi / T_{\rm max}) \cdot \exp(2)$ if $m_\chi > T_{\rm
max}$, since this contribution needs two factors of the ``soft''
particle density. The exact form of this exponential cut--off is
debatable, but it is clear that the soft--soft contribution should
decrease exponentially, and hence quickly become irrelevant, once
$m_\chi > T_{\rm max}$.  In the present case, the soft--soft
contribution is overwhelmed by the hard--soft one even prior to this
sharp decrease (which makes the exact form of the cut--off
irrelevant). This is because $T_0 < T_{\rm R}$ as long as $m_\chi <
1.1 \cdot 10^{10}$ GeV, in which case hard--soft scatterings which
occur at the very last stage of inflaton decay are the dominant source
of $\chi$ production. However, the hard--soft contribution itself
drops very quickly for $m_\chi > 10^{11}$ GeV, since $T_0$ exceeds
$T_{\rm max}$ in this mass range. We have multiplied
eq.(\ref{hsrate4}) by $\exp(-T_0/T_{\rm max}) \cdot \exp(1)$ if $T_0 >
T_{\rm max}$. Once again the exact form of this cut--off is not very
important, since the hard--hard contribution becomes dominant above
this point until it is kinematically suppressed, and then forbidden,
for $m_\chi \approx 5 \cdot 10^{12}$ GeV. It is interesting to note
that hard--hard scatterings can very efficiently produce
``wimpzillas'' up to this kinematical cut-off. In fact, for this
choice of $T_{\rm R}$, $m_\phi$ and $\alpha_\chi$ the overclosure
bound (for a strictly stable $\chi$) and the astrophysical bound (for
a late decaying $\chi$) require that $m_\chi$ must be bigger than
$m_\phi/2$, otherwise hard--hard scatterings will produce $\chi$ in
unacceptably large abundances. Recall, however, eq.(\ref{tmax}) and
eq.(\ref{hhrate}) which show that $\Omega_\chi^{\rm hh} \propto
T^4_{\rm R}$ for fixed $m_\phi$; reducing the re--heat temperature by
a factor of 2 would therefore result in an acceptable $\chi$ relic
density if $T_0 > T_{\rm max}$. In contrast, even without the
exponential cut--off, the soft--soft contribution would have satisfied
the overclosure bound for $m_\chi > 5 \cdot 10^{10}$ GeV, i.e. for the
parameters of Fig.~2a the hard--hard contribution raises the bound on
$m_\chi$ by two orders of magnitude.

The lower reheat temperature in Fig.~2b leads to a lower value of
$T_{\rm max} \simeq 6 \cdot 10^7$ GeV.  In this case soft--soft
production of $\chi$ pairs actually were in equilibrium for $m_\chi <
2 \cdot 10^7$ GeV, but this contribution again falls very sharply once
$m_\chi > T_{\rm max}$, where the hard--soft contribution becomes
dominant.  A kink in the hard--soft contribution appears at $m_\chi
\simeq 4 \cdot 10^8$ GeV, where $T_0$ starts exceeding $T_{\rm
R}$\footnote{This kink is also barely visible in Fig.~2a at $m_\chi
\simeq 10^{10}$ GeV.}. The curve flattens out just before the kink
since here the $2 \rightarrow 2$ contribution to the production cross
section become important.  The hard--soft contribution decays rapidly
for $m_\chi > 8 \cdot 10^9$ GeV, where $T_0 > T_{\rm max}$. Then the
hard--hard contribution naturally takes over and quickly becomes the
only source until the kinematical cut--off at $5 \cdot 10^{12}$
GeV. The overclosure bound now requires $m_\chi \geq 10^8$ GeV (by a
funny coincidence the soft--soft and hard--soft contributions are
comparable around this lower limit). On the other hand, for an
unstable $\chi$ the astrophysical bound in eq.(\ref{decbound})
requires $T_0 > T_{\rm max}$, i.e. $m_\chi > 10^{10}$ GeV, if
$\tau_\chi \sim \tau_{\rm U}$ (hard--soft and hard--hard contributions
are again coincidentally comparable around this limit). Notice that the
hard--soft contribution have increased this limit by about a factor of
10. It is also observed that now the hard--hard contribution does not
suffice to make $\chi$ an interesting Dark Matter particle for any
value of $m_\chi$. This again follows from the $T_{\rm R}^4$ behavior
of this contribution.

Finally, Fig.~2c displays the results for $T_{\rm R} = 3$ MeV, near
the lower limit required for successful nucleosynthesis; we also chose
a smaller inflaton mass $m_\phi = 10^8$ GeV so that this very low
reheat temperature will be achieved more naturally. In this case
$T_{\rm max} \simeq 600$ GeV. A main difference from the previous
cases is that now the soft--soft contribution is very small even for
masses as low as $m_\chi \simeq 1$ GeV. The reason is that the
abundance of $\chi$ particles, produced from soft--soft scatterings,
is determined by their annihilation which is at equilibrium in this
case if $m_\chi \leq 200$ GeV. As a result the soft--soft contribution
is subdominant for almost all values of $m_\chi$; as usual, it cuts
off sharply once $m_\chi > T_{\rm max}$. Again a kink in the curve
depicting the hard--soft contribution is recognized at $m_\chi \simeq
200$ GeV, above which $T_0 > T_{\rm R}$. Even the hard--soft
contribution essentially vanishes for $m_\chi > 10^5$ GeV, where $T_0$
exceeds $T_{\rm max}$. The hard--hard contribution is now completely
negligible since $T_{\rm R}$ is very small. The only meaningful
constraint in this case is that from eq.(\ref{champsbound}). It is
seen that the hard--soft contribution to $\chi$ production again
increases the lower bound on $m_\chi$ by about one order of magnitude,
to about $10^4$ GeV, compared to the bound derived in Ref.\cite{ky},
which only includes the soft--soft contribution. Naturalness arguments
indicate that the mass of the lightest superparticle (LSP) should not
exceed 1 TeV; this argument is independent of whether or not the LSP
is charged. The constraint of Fig.~1c is not compatible with this
bound. Indeed, we find that once the hard--soft contribution is
included, even for $T_{\rm R} = 1$ MeV, $m_\chi = 1$ TeV is compatible
with the bound (\ref{champsbound}) only if $m_\phi \leq 5 \cdot 10^5$
GeV. This indicates that scenarios with a charged LSP would require a
model of low scale inflation.

\setcounter{footnote}{0} 
One comment is in order before moving on to the next subsection. The
production of $\chi$ particles from soft--soft scatterings is
independent of the main channel through which inflaton decays, so long
as the decay products thermalize sufficiently rapidly (i.e. such that
$m_\chi < T_{\rm max}$). For example, assume that (for some reason)
the inflaton mostly decays to leptons while $\chi$ can only be
produced from the scattering of quarks. In this case $\Omega^{\rm
ss}_{\chi}$ remains essentially unchanged since electroweak gauge
interactions bring quarks into equilibrium with a thermal bath created
by leptons. On the other hand, $\Omega^{\rm hs}_{\chi}$ and
$\Omega^{\rm hh}_{\chi}$ will be suppressed since a hard lepton cannot
produce $\chi$ from $2 \rightarrow 2$ scattering off soft particles in
the thermal bath and/or off another hard lepton. In this scenario
$\chi$ production from hard--soft or hard--hard processes can still
occur via $2 \rightarrow 4$ scattering reactions; if $q \bar q
\rightarrow \chi \bar \chi$ is allowed\footnote{$\chi = \bar \chi$ if
$\chi$ is a Majorana particle.}, so are $\ell q \rightarrow \ell q
\bar \chi \chi$ and $\ell^+ \ell^- \rightarrow \bar q q \bar \chi
\chi$. The corresponding contributions $\Omega^{\rm hs}_{\chi}$ and
$\Omega^{\rm hh}_{\chi}$ will then be smaller than our estimates in
eqs.(\ref{hsrate3}), (\ref{hsrate4}) and (\ref{hhrate}) by several
orders of magnitude. This possibility may be relevant to the
production of exotic electroweak gauge singlet particles with large
masses, thus relaxing constraints on the model parameters in that
case. However, the estimates for the LSP and charged stable particle
examples will remain unaffected since these species can be produced
from $2 \rightarrow 2$ scatterings of all standard model particles.

\subsection{Particle production from inflaton decay}

We now discuss the direct production of $\chi$ particles in inflaton
decay whose importance has recently been noticed \cite{ad1}. Most
inflatons decay at $T \simeq T_{\rm R}$; moreover, the density of
$\chi$ particles produced in earlier inflaton decays will be greatly
diluted. Since inflaton decay conserves energy, the density of
inflatons can be estimated as $n_\phi \simeq 0.3 g_* T_R^4 /
m_\phi$. Let us denote the branching ratio for $\phi \rightarrow \chi$
decays (more accurately, the average number of $\chi$ particles
produced in each $\phi$ decay) by $B(\phi \rightarrow \chi)$. The
$\chi$ density from $\phi$ decay can then be estimated as
\cite{rosenfeld}: 
\beq \label{phidec}
\Omega^{\rm decay}_\chi h^2 \simeq 2 \cdot 10^8 B(\phi \rightarrow
\chi) \frac {m_\chi} {m_\phi} \frac {T_{\rm R}} {1 \ {\rm GeV}}.
\eeq
Eq.(\ref{phidec}) holds if the $\chi$ annihilation rate is smaller
than the Hubble expansion rate at $T \simeq T_{\rm R}$, which requires
\beq \label{nonequil}
\frac{m_\phi} {M_{\rm Planck}} > 5 B(\phi \rightarrow \chi)
\alpha_\chi^2 \left( \frac {T_R} {m_\chi} \right)^2 \left( \frac {g_*}
{200} \right)^{1/2}.
\eeq
This condition will be satisfied in chaotic inflation models with
$m_\phi \sim 10^{-5} M_{\rm Planck}$, if $m_\chi$ is large enough to avoid
overclosure from thermal $\chi$ production alone. It might be violated
in models with a light inflaton. In that case the true $\chi$ density at
$T_{\rm R}$ can be estimated by equating the annihilation rate with
the expansion rate:
\beq \label{ommax}
\Omega_\chi^{\rm max} \simeq \frac {5 \cdot 10^7} {\alpha_\chi^2}
\frac {m_\chi^3} {(1 \ {\rm GeV}) \cdot M_{\rm Planck} T_{\rm R}}
\left( \frac {200} {g_*} \right)^{1/2}.
\eeq
This maximal density violates the overclosure constraint $\Omega_\chi
< 1$ badly for the kind of weakly interacting ($\alpha_\chi \lsim
0.1$), massive ($m_\chi \gg T_{\rm R}$ and $m_{\chi} \gsim 1$ TeV)
particles we are interested in.\footnote{Eq.(\ref{ommax}) describes
the maximal $\chi$ density if $\chi$ decouples at $T \sim T_R$. It is
not applicable to WIMPs decoupling at $T < T_{\rm R}$.}  For the
remainder of this article we will therefore estimate the $\chi$
density from inflaton decay using eq.(\ref{phidec}).

We now discuss estimates\footnote{In Ref.\cite{rosenfeld} simultaneous
production of $\chi$ from the thermal bath and from direct inflaton
decay has been studied and a bound on the branching ratio for such a
decay mode is derived. Our aim here is to show that higher order
processes naturally provide a decay mode, even if $\chi$ is not
directly coupled to the inflaton.} of $B(\phi \rightarrow \chi)$. This
quantity is obviously model dependent, so we have to investigate
several scenarios. The first, important special case is where $\chi$
is the LSP. If $m_\phi$ is large compared to typical visible--sector
superparticle masses, $\phi$ will decay into particles and
superparticles with approximately equal probability\footnote{The
non-thermal production of LSP has also been considered in other
contexts. For a neutralino LSP these include its production from the
decay of Q-balls \cite{qball}, in cosmic string decays as a possible
solution to the observational conflicts of WIMP cold dark matter
\cite{bran}, in decay of the Polonyi field \cite{kmy}, and in moduli
decay in connection with models of anomaly--mediated supersymmetry
breaking \cite{mr} and models with intermediate scale unification
\cite{kmt}. Non-thermal production of axino LSP, as Dark Matter, from
neutralino decay has been considered in \cite{axino}.}. This can be
illustrated for two possible cases: when the main inflaton decay mode
is via a superpotential coupling, and for a gravitationally decaying
inflaton. In the first case consider the inflaton superfield $\Phi$,
comprising the inflaton $\phi$ and its superpartner inflatino $\tilde
\phi$, and a chiral supermultiplet $\Psi$, which comprises a complex
scalar field $\tilde \psi$ and its fermionic partner $\psi$, with the
superpotential coupling
\beq
W \supset {1 \over 2} m_\phi \Phi^2 + {1 \over 2} h_\chi \Phi {\Psi}^2.
\eeq
This superpotential generates the following terms
\beq
h_\chi \phi \bar \psi \psi~~;~~h_\chi m_\phi \phi^* \tilde\psi
\tilde\psi, 
\eeq
in the Lagrangian. It is easily verified that $\phi$ decays to both of
$\psi$ and $\tilde \psi$ (which in turn decay to matter particles
plus the LSP) at a rate given by $\Gamma = (h^2_\chi/8 \pi) m_\phi$.

\setcounter{footnote}{0}
Now let us turn to the case of a gravitationally decaying inflaton. As
a simple example, consider minimal supergravity where the scalar
potential is given by \cite{nilles}
\beq \label{gdecay}
V = e^G \left ({\partial G \over \partial {\varphi}_i} {\partial G
\over \partial {\varphi}^{*}_{i}} - {3 \over M^{2}_{\rm Planck}}
\right ) M^6_{\rm Planck}.
\eeq
Here $G$ is the K\"ahler function defined by
\beq
G = {{\varphi_i} \varphi^{*}_{i} \over M^{2}_{\rm Planck}} + 
\log \left ({{|W|}^{2} \over M^{6}_{\rm Planck}} \right ),
\eeq
where $\varphi_i$ are the scalar fields in the theory. The inflation
sector superpotential looks like $W \sim (1/2) m_\phi {(\Phi - v)}^2$
around the minimum of the potential where $v$ denotes the inflaton VEV
at the minimum\footnote{In the absence of any superpotential coupling
to other multiplets which provides a linear term in $\phi$, a non-zero
$v$ is required for the inflaton decay to take place.}. The
superpotential also includes the familiar Yukawa couplings for the
matter sector, e.g. $h_u {\rm H}_u {\rm Q} {\rm u}$ where $\rm Q$ and
$\rm u$ denote the superfields containing the doublet of left-handed
and the singlet of right-handed (s)quarks, respectively, and ${\rm
H}_u$ is the superfield which contains the Higgs doublet giving mass
to the up-type quarks. Then it is easy to see that Eq.(\ref{gdecay})
leads to the following term (among others)
\beq \label{scalcoup}
h_u {v m_\phi \over M^{2}_{\rm Planck}} \phi^* H_u {\tilde Q} {\tilde u}
\eeq
in the Lagrangian for inflaton decay to three scalars, including two
superparticles\footnote{The main mode for decay to three scalars is to
$H_u$ and two squarks of the third generation, because of the large
top Yukawa coupling.}. This implies that the rate for production of
particles and sparticles through inflaton decays into three light
scalars is approximately the same. 

Once one goes beyond minimal supergravity, the inflaton can also decay
to gauge fields and gauginos. Consider the case where the gauge
superfields have nonminimal kinetic terms, as a result of $M_{\rm
Planck}$ suppressed couplings to the inflaton, in the following form
\beq \label{nonminimal}
f_{\alpha \beta} = \left[1 + a \left({\phi \over M_{\rm
Planck}}\right)+ b \left({\phi \over
M_{\rm Planck}}\right)^2 + ....\right]~ \delta_{\alpha \beta},
\eeq
Then to the leading order one finds the term
\beq \label{gauge}
a {\phi \over M_{\rm Planck}} F^{\alpha}_{\mu \nu} F^{\mu \nu, \alpha},
\eeq
for the inflaton coupling to gauge fields, where $\alpha$ represents
the relevant gauge group index. This results in inflaton decay to a
pair of gauge quanta at the rate $\Gamma \sim a^2
m^{3}_{\phi}/M^{2}_{\rm Planck}$. The corresponding term from the
kinetic energy of the gauginos results in a derivative coupling and
hence a decay rate which is suppressed by a factor $m^{2}_{\tilde
g}/m^{2}_{\phi}$, where $m_{\tilde g}$ is the gaugino mass
\cite{mr}. However, there exists another term in the Lagrangian
\cite{nilles} which
is responsible for gaugino mass from supersymmetry breaking by the
inflaton energy density:
\beq \label{gaugino}
F_{\phi} {\partial f^{*}_{\alpha \beta} \over \partial \phi^*}
\lambda^{\alpha} \lambda^{\beta},
\eeq
where $F_\phi$ is the $F$-term associated with the inflaton
superfield, given by $\simeq m_\phi \phi$. It is easy to see
that the term in (\ref{gaugino}), to the leading order, results in
inflaton decay to two gauginos at the rate $\Gamma \sim a^2
m^3_\phi/M^2_{\rm Planck}$, which is comparable to that for inflaton
decay to two gauge quanta\footnote{Note that moduli decay to gauginos
through the term in (\ref{gaugino}) will be suppressed \cite{mr}. The
reason is
that the superpotential is (at most) linear in the moduli superfield,
so that the corresponding $F-$term does not contain the moduli
field. In contrast, in almost all inflationary models the dependence
of the superpotential on the inflaton superfield is quadratic or
higher.}.

Moreover, all superparticles will quickly decay into the LSP and some
standard particle(s). As long as $m_\chi > T_{\rm R}$, the time scale
for these decays will be shorter than the superparticle annihilation
time scale even if $\alpha_\chi \simeq 0.1$. As a result, if $\chi$ is
the LSP, then $B(\phi \rightarrow \chi) \simeq 1$, independently of
the nature of the LSP.

Another possibility is that the inflaton couples to all particles with
more or less equal strength, e.g. through non--renormalizable
interactions. In that case one expects $B(\phi \rightarrow \chi) \sim
1/g_* \sim 1/200$. However, even if $\phi$ has no direct couplings to
$\chi$, the rate (\ref{phidec}) can be large. The key observation is
that $\chi$ can be produced in $\phi$ decays that occur in higher
order in perturbation theory whenever $\chi$ can be produced from
annihilation of particles in the thermal plasma. In most realistic
cases, $\phi \rightarrow f \bar f \chi \bar \chi$ decays will be
possible if $\chi$ has gauge interactions, where $f$ stands for some
gauge non--singlet with tree--level coupling to $\phi$. A diagram
contributing to this decay is shown in Fig.~3. Note that the part of
the diagram describing $\chi \bar \chi$ production is identical to the
diagram describing $\chi \bar \chi \leftrightarrow f \bar f$
transitions. This leads to the following estimate:
\beq \label{fourbody}
B(\phi \rightarrow \chi)_4 \sim \frac {C_4 \alpha_\chi^2} {96 \pi^3}
\left( 1 - \frac {4 m_\chi^2} {m_\phi^2} \right)^2 \left( 1 - \frac {2
m_\chi} {m_\phi} \right)^{5 \over 2},
\eeq
where $C_4$ is a multiplicity (color) factor. The phase space factors
have been written in a fashion that reproduces the correct behavior
for $m_\chi \rightarrow m_\phi/2$ as well as for $m_\chi \rightarrow
0$. This estimate provides a lower bound on $B(\phi \rightarrow \chi)$
under the conditions assumed for our calculation of $\Omega_\chi^{\rm
hs}$ and $\Omega_\chi^{\rm hh}$; whenever a primary inflaton decay
product can interact with a particle in the thermal plasma, or with
another primary decay product, to produce a $\chi \bar \chi$ pair, $\phi
\rightarrow \chi$ four--body decays must exist. It is easy to see that
the contribution (\ref{phidec}) to the $\chi$ density from inflaton
decay will exceed the hard--hard contribution (\ref{hhrate}), if
$T_{\rm max}$ can be estimated from eq.(\ref{tmax}).  If $T_0 > T_{\rm
R}$, the decay contribution will exceed the hard--soft contribution if
$m^6_\chi > 0.3 m^2_\phi T_{\rm R}^4 / \alpha^3$, which is true in
almost all models that avoid overclosure from thermal $\chi$
production alone. Even if $T_0 < T_{\rm R}$, the decay contribution
will dominate over the hard--soft contribution if
\beq \label{compare}
\alpha^2 > 150 \left[ \left( \frac {T_{\rm R}} {m_\chi}
\right)^2 + \frac {T_R} {\alpha m_\phi} \right],
\eeq
where the first and second term in the square bracket describe $\chi$
pair production from $2 \rightarrow 3$ and $2 \rightarrow 2$
processes, respectively; see eq.(\ref{hsprod}). This condition can be
mildly violated, i.e. in some cases the hard--soft contribution may
exceed the decay contribution. For example, in the scenarios
considered in Figs.~2, as long as $m_\chi \ll m_\phi$ we find
$\Omega_\chi^{\rm decay} \sim 6 C_4 \cdot 10^5 m_\chi / (10^{10} \
{\rm GeV})$ in a); $\sim 0.6 C_4 m_\chi / (10^7 \ {\rm GeV})$ in b);
and $\sim 2 \cdot 10^{-10} C_4 m_\chi / {\rm GeV}$ in c). For $C_4 =
1$, there is a narrow range of $m_\chi$ in Figs.~2a,b where
$\Omega_\chi^{\rm decay} < \Omega^{\rm hs}_\chi$; however, in this
range $\Omega_\chi$ exceeds the upper bound of 1 significantly. We
thus conclude that the decay contribution to $\Omega_\chi$ will
usually dominate over nonthermal $\chi$ production from inflaton decay
products if four--body $\phi \rightarrow \chi$ decays exist.

\vspace*{6mm}
\begin{center}
\SetScale{0.6} \SetOffset(40,40)
\begin{picture}(225,125)(0,0)
\DashLine(0,50)(75,50){5} \Text(0,25)[l]{$\phi$}
\Vertex(75,50){3}
\ArrowLine(125,25)(75,50) \Text(80,16)[h]{$\bar{f}$}
\ArrowLine(75,50)(125,75) \Text(60,50)[t]{$f$}
\Vertex(125,75){3}
\Photon(125,75)(175,100){5}{4} 
\Vertex(175,100){3}
\ArrowLine(125,75)(175,50) \Text(110,30)[h]{$f$}
\ArrowLine(175,100)(225,75) \Text(145,45)[r]{$\chi$}
\ArrowLine(225,125)(175,100) \Text(145,75)[r]{$\bar \chi$}              
\end{picture}
\vspace*{-13mm}

\noindent
{\bf Fig. 3:}~Sample diagram for $\chi$ production in four-body
inflaton decay. 
\end{center}
\vspace*{3mm}

Occasionally one has to go to even higher order in perturbation
theory to produce $\chi$ particles from $\phi$ decays. For example, if
$\chi$ has only strong interactions but $\phi$ only couples to $SU(3)$
singlets, $\chi \bar \chi$ pairs can only be produced in six body
final states, $\phi \rightarrow f \bar f q \bar q \chi \bar \chi$. A
representative diagram can be obtained from the one shown in Fig.~3 by
replacing the $\chi$ lines by quark lines, attaching an additional
virtual gluon to one of the quarks which finally splits into $\chi
\bar \chi$. The branching ratio for such six body decays can be
estimated as
\beq \label{sixbody}
B(\phi \rightarrow \chi)_6 \sim \frac {C_6 \alpha_\chi^2 \alpha^2}
{1.1 \cdot 10^7}
\left( 1 - \frac {4 m_\chi^2} {m_\phi^2} \right)^4 \left( 1 - \frac {2
m_\chi} {m_\phi} \right)^{9 \over 2}.
\eeq
Another example where $\chi \bar \chi$ pairs can only be produced in
$\phi$ decays into six body final states occurs if the inflaton only
couples to fields that are singlets under the SM gauge group,
e.g. right--handed (s)neutrinos $\nu_R$ \cite{ahky}. These
(s)neutrinos can emit a virtual Higgs boson, which can split into a
top quark-antiquark pair; one of which can emit a virtual gluon, which
in turn splits into a (strongly interacting) $\chi \bar \chi$ pair. In
this scenario the factor $\alpha^2$ in eq.(\ref{sixbody}) would have
to be replaced by the combination of Yukawa couplings $h_{\nu_R}^2
h_t^2 / (16 \pi^2)$. If $2m_\chi < m_{\nu_R}$, $\chi \bar \chi$ pairs
can already be produced in four body final states from $\nu_R$
decay. The effective $\phi \rightarrow \chi$ branching ratio would
then again be given by eq.(\ref{fourbody}), with $m_\phi$ replaced by
$m_{\nu_R}$ in the kinematical factors.

Finally, in supergravity models with explicit (supersymmetric) $\chi$
mass term there in general exists a coupling between $\phi$ and either
$\chi$ itself or, for fermionic $\chi$, to its scalar superpartner, of
the form $a \left (m_\phi m_\chi/M_{\rm Planck}\right ) \phi^* \chi \chi
+ {\rm h.c.}$ in the scalar potential\footnote{This term also induces
an $A$-term from supersymmetry breaking by the inflaton energy
density.}; this term is completely
analogous to the one shown in (\ref{scalcoup})\footnote{Note that a
superpotential mass term for the $\chi$
multiplet will be allowed under a $Z_2$
discrete symmetry. For a continuous symmetry one requires two
multiplets $\chi_1$
and $\chi_2$ with opposite charges.}. A reasonable estimate
for the coupling strength is
\cite{aem} $a \sim v/M_{\rm Planck}$, unless an $R-$symmetry
suppresses $a$. Assuming that most inflatons decay into other
channels, so that $\Gamma_{\rm decay} \sim \sqrt{g_*} T_{\rm
R}^2/M_{\rm Planck}$ remains valid, this gives
\beq \label{gravbr}
B(\phi \rightarrow \chi) \sim \frac {a^2 m^2_\chi m_\phi} {16 \pi
\sqrt{g_*} M_{\rm Planck} T^2_{\rm R} } \left( 1 - \frac {4 m_\chi^2}
{m_\phi^2} \right)^{1 \over 2}.
\eeq

The production of $\chi$ particles from inflaton decay will be
important for large $m_\chi$ and large ratio $m_\chi / T_{\rm R}$, but
tends to become less relevant for large ratio $m_\phi / m_\chi$. Even
if $m_\chi < T_{\rm max}$, $\chi$ production from the thermal plasma
(\ref{ssrate1}) will be subdominant if
\beq \label{compare1}
\frac {B(\phi \rightarrow \chi)} {\alpha_\chi^2} > \left( \frac {100
T_{\rm R}} {m_\chi} \right)^6 \frac {m_\phi} {m_\chi} \frac {1 \ {\rm
TeV}} {m_\chi}.
\eeq
The first factor on the r.h.s. of (\ref{compare1}) must be $\lsim
10^{-6}$ in order to avoid over--production of $\chi$ from thermal
sources alone. Even if $\phi \rightarrow \chi$ decays only occur in
higher orders of perturbation theory, the l.h.s. of (\ref{compare1})
will be of order $10^{-4}$ ($10^{-10}$) for four (six) body final
states, see eqs.(\ref{fourbody}), (\ref{sixbody}); if two--body
tree--level $\phi \rightarrow \chi \bar \chi$ decays are allowed, the
l.h.s. of (\ref{compare1}) will usually be bigger than unity. We thus
see that even for $m_\phi \sim 10^{13}$ GeV, as in chaotic inflation
models, and for $m_\chi \simeq 10^3 T_{\rm R}$, $\chi$ production from
decay will dominate if $m_\chi \gsim 10^7 \ (10^{10})$ GeV for four
(six) body final states; this agrees with the numerical results shown
in Fig.~2. As a second example, consider LSP production in models with
very low reheat temperature. Naturalness arguments imply that the LSP
mass should lie within a factor of five or so of 200 GeV. Recall that
in this case $B(\phi \rightarrow \chi) = 1$. Taking $\alpha_\chi \sim
0.01$, we see that $\chi$ production from decay will dominate over
production from the thermal plasma if $m_\phi < 6 \cdot 10^7$ GeV for
$T_{\rm R} = 1$ GeV; this statement will be true for all $m_\phi \lsim
10^{13}$ GeV if $T_{\rm R} \lsim 100$ MeV.

In \cite{ad1} we showed that the decay contribution (\ref{phidec}) by
itself leads to very stringent constraints on models with massive
stable $\chi$ particles. In particular, charged stable particles with
mass below $\sim 100$ TeV seem to be excluded, unless $m_\chi >
m_\phi/2$. In case of a (neutral) LSP with mass around 200 GeV, the
overclosure constraint implies $m_\phi / T_{\rm R} > 4 \cdot 10^{10}$,
i.e. a very low reheat temperature, unless $\chi$ was in thermal
equilibrium below $T_R$; recall that $B(\phi \rightarrow \chi) = 1$ in
this case. Finally, if $m_\phi \sim 10^{13}$ GeV a ``wimpzilla'' with
mass $m_\chi \sim 10^{12}$ GeV will be a good Dark Matter candidate
only if it has a very low branching ratio, $B(\phi \rightarrow \chi)
\sim 5 \cdot 10^{-8} \ {\rm GeV} / T_{\rm R}$, i.e. if its couplings
to ordinary matter are very small.

Our calculation is also applicable to entropy--producing particle
decays that might occur at very late times. If $\chi$ is
lighter than this additional $\phi'$ particle all our
expressions go through with the obvious replacement $\phi \rightarrow
\phi'$ everywhere. More generally our result holds if $\phi$ decays
result in a radiation-dominated era with $T_{\rm R} > m_{\phi'}$. If
$\phi'$ is sufficiently long--lived, the Universe will eventually
enter a second matter--dominated epoch. $\phi'$ decays then give rise
to a second epoch of reheating, leading to a radiation--dominated
Universe with final reheating temperature $T_{\rm R_f}$, and
increasing the entropy by a factor $m_{\phi'} / T_{\rm R_f}$. This
could be incorporated into eq.(\ref{phidec}) by replacing $T_{\rm R}
\rightarrow T_{\rm R} T_{\rm R_f} / m_{\phi'} > T_{\rm R_f}$.

In a similar vein, consider the production of a neutral LSP from
gravitino decay in gravity-mediated models of supersymmetry breaking;
note that in this case the decaying particle does not dominate the
energy density of the Universe. Gravitinos with mass $m_{3/2} \sim
100~{\rm GeV}-3$ TeV have a lifetime $\tau > 1$ s, which can ruin the
success of nucleosynthesis if gravitinos are produced in large
abundances \cite{bbn}. This, as well known, leads to constraints on
the reheating temperature of the Universe $T_{\rm R}$
\cite{gravitino}. On the other hand, for $m_{3/2} > 10$ TeV the
gravitino decays before nucleosynthesis and has no effect on the light
element abundances\footnote{A gravitino mass which is much larger than
superparticle masses can be naturally found, for example, in models of
no-scale supergravity \cite{noscale}.}. However, even in this case a
significant upper limit on $T_{\rm R}$ can be derived from the
following argument. If the LSP has a mass $\geq 100$ GeV, gravitinos
(as heavy as $10^6$ GeV) decay much after the freeze--out of LSP
annihilation. The overclosure bound then results in a constraint on
the gravitino number to entropy ratio, $n_{3/2}/s \leq 4 \times
10^{-11}$, even when the gravitinos decay before the onset of
nucleosynthesis. For thermal gravitino production, where $n_{3/2}/s
\simeq 10^{-11} (T_{\rm R}/10^{11}~{\rm GeV})$ \cite{gravitino}, this
results in the limit $T_{\rm R} \leq 10^{11}$ GeV; including possible
non--thermal production of gravitinos \cite{kallosh} will presumably
sharpen this limit.

Finally, let us point out possible implications of deviating from two
major assumptions which we made throughout this article: a significant
branching ratio of primary inflaton decays into known SM particles
(possibly including their superpartners), and allowed
(pair--)production of $\chi$ particles in the scattering of matter
particles. First, consider the case when the inflaton exclusively
decays to exotic light particles while $\chi$ is produced through its
coupling to matter particles. Assume that these exotic particles only
couple with strength $\alpha' \ll \alpha$ to SM particles. This
results in a smaller $T_{\rm max}$; depending on the details of the
model, one has to replace the factor $\alpha^3$ in eq.(\ref{tmax})
either by $\alpha \alpha'^2$ or by $\alpha'^2$. We consider the latter
possibility and assume for simplicity that $T_{\rm R}$ remains
unchanged. The thermal contribution $\Omega^{\rm ss}_{\chi}$ will be
significantly reduced if the new value of $T_{\rm max}$ is below
$m_\chi$ by more than one order of magnitude or so. We therefore
require that $T_{\rm max} \simeq m_\chi$. Let us find $\Omega^{\rm
decay}_{\chi}$ by assuming that $\Omega^{\rm ss}_{\chi} \simeq 1$. The
main difference is that $\chi$ production from inflaton decay now
occurs through six body final state diagrams. with $\alpha^2$ in
eq.(\ref{sixbody}) replaced by $\alpha'^2$. The requirement that
$T_{\rm max} \geq m_\chi$ results in a bound on $\alpha'^2$, see
eq.(\ref{tmax}). For $m_\chi = 2\cdot 10^3 \alpha_\chi^{2/7} T_{\rm
R}$, which saturates $\Omega^{\rm ss}_{\chi}$, and after using the
bound on $\alpha'^2$ in (\ref{phidec}), we find
\beq \label{newsix1}
\Omega^{\rm decay}_{\chi} \gsim 5 \cdot 10^3\, \alpha_\chi^{32/21}
{m^{5/3}_\chi \over m^{2/3}_\phi M_{\rm Planck}} {m_\chi \over
1~{\rm GeV}}, 
\eeq
for $\chi$ production in six body decay of the inflaton. Even for the
most conservative choice, $m_\phi \simeq 10^{13}$ GeV,
eq.(\ref{newsix1}) requires $m_\chi < 7 \cdot 10^9$ GeV for
$\alpha_\chi \sim 0.01$. Note that $\chi$ production from hard--soft
and hard--hard scatterings also only occurs through higher order
processes, thus $\Omega^{\rm hs}_{\chi}$ and $\Omega^{\rm hh}_{\chi}$
are suppressed by a few orders of magnitude compared to expressions
(\ref{hsrate3}), (\ref{hsrate4}) and (\ref{hhrate}). Moreover, $T_{\rm
max} \ll m_\chi$ if $\alpha'$ is very small, implying that
$\Omega^{\rm ss}_{\chi}$ will be exponentially suppressed. In this
case $\chi$ production from inflaton decay will dominate, since it is
only suppressed by a factor $\alpha'^2$.

Second, consider the case where the inflaton decays to matter but
$\chi$ particles can only be produced from scatterings of some
intermediate particle $\chi'$ whose interactions with matter has a
strength $\alpha' \ll \alpha$. If $\alpha'$ is not very small $\chi'$
will be in thermal equilibrium with matter and, provided $T_{\rm max}
> m_\chi$, $\Omega^{\rm ss}_{\chi}$ will only be reduced by a
statistics factor, since now only a small fraction of all soft--soft
scatters can produce $\chi$ particles. On the other hand, $\Omega^{\rm
hs}_{\chi}$ and $\Omega^{\rm hh}_{\chi}$ will decrease much more. The
hard--soft and hard--hard scatterings now can produce $\chi$'s only in
four--body final states, e.g. $\bar f f \rightarrow \bar \chi' \chi'
\bar \chi \chi$, with cross section $\propto \alpha_\chi^2
\alpha'^2$. By using the expression in (\ref{temp}) a bound on
$\alpha'$ is found in order for $\chi'$ to be at thermal equilibrium
for $T \geq m_\chi$, which requires $\alpha'^2 m_\chi \geq H$. If we
now assume $\Omega^{\rm ss}_{\chi} \simeq 1$, taking $g_*=200$, we
have 
\beq \label{newsix2} 
\Omega^{\rm decay}_{\chi} \gsim 10^5 \alpha_\chi^{12/7} {m^{2}_{\chi}
\over m_\phi M_{\rm Planck}} {m_\chi \over 1~{\rm GeV}}, 
\eeq
from six body decay of the inflaton. Again taking $\alpha_\chi = 0.01$
and $m_\phi = 10^{13}$ GeV, this results in the bound $m_\chi \leq
10^{10}$ GeV in order not to overclose the Universe. As a closing
remark, we shall notice that such a rather contrived scenario might be
realized for exotic $\chi$ particles (e.g. ``wimpzillas''), but not
when $\chi$ is the LSP or a charged stable particle, which have
electroweak gauge couplings to ordinary matter.

\section{Summary and Conclusions}

In this article we studied the thermalization of perturbative inflaton
decay products, with emphasis on applications to the production of
massive stable or long--lived particles $\chi$. We found that a
thermal plasma should form well before inflaton decay is complete, if
the theory contains light or massless gauge bosons; $2 \rightarrow 3$
reactions where a fairly energetic, nearly collinear particle is
emitted in the initial or final state play a crucial role here. The
existence of light gauge bosons is required since only gauge boson
exchange in the $t-$ or $u-$channel leads to cross sections that
significantly exceed $\alpha^2 / m_\phi^2$, where $\alpha$ is the
relevant (gauge) coupling strength and $m_\phi$ the inflaton
mass. This indicates that reheating might be delayed greatly if some
scalar field(s) break all gauge symmetries during this epoch, which
may naturally happen in the presence of supersymmetric flat
directions.

Even if massless gauge bosons exist, thermalization takes a finite
amount of time. As a result, the maximal temperature of the thermal
plasma will usually be well below $m_\phi$ (but can exceed the reheat
temperature $T_{\rm R}$ significantly); this limits the region of
parameter space where {\em thermal} $\chi$ production can play a
role. On the other hand, it allows the very energetic primary inflaton
decay products to produce $\chi$ particles either in collisions with
the thermal plasma (``hard--soft'' scattering), or with each other
(``hard--hard'' reactions). We estimated the rate for these reactions
in the simple approximation where the primary inflaton decay products
have energy $m_\phi / 2$ for one thermalization time, and then have
energy $T$. If $m_\phi$ is rather close to $m_\chi$ this will probably
overestimate the true rate, since then the energy of the primary decay
products will drop below the production threshold faster than in our
approximation. On the other hand, if $m_\phi \gg m_\chi$, our
approximation will likely be an underestimate, since in the process of
thermalization a single particle with $E \sim m_\phi/2$ can produce
several (most likely nearly collinear) particles with $m_\phi/2 > E >
m_\chi$, all of which can contribute to nonthermal $\chi$ production;
note that this also allows $\chi$ production from the scattering of
``hard'' particles even if the primary inflaton decay products do not
couple directly to $\chi$. We found that the hard--soft contribution
will dominate over the hard--hard one if it is still kinematically
allowed at $T = T_{\rm R}$, but can otherwise be subdominant; either
of these two new production mechanisms can exceed the rate from purely
thermal $\chi$ production.

We also discussed $\chi$ production in inflaton decay. We pointed out
that decays of this kind must be allowed, at least in four--body final
states, if $\chi$ particles can be produced in collisions of primary
inflaton decay products; in certain (somewhat contrived) scenarios one
may have to consider six--body final states. In fact, in most cases
this seems to be the most important nonthermal (but perturbative)
mechanism producing massive particles with $m_\chi < m_\phi/2$; this
contribution also often exceeds thermal $\chi$ production by several
orders of magnitude, even if $m_\chi$ is below the maximal temperature
of the thermal plasma. Nonthermal $\chi$ production therefore
significantly sharpens limits on model parameters that follow from
upper bounds on the $\chi$ relic density. For example, if $m_\phi$ is
well above visible sector superparticle masses, each inflaton decay
will produce ${\cal O}(1)$ lightest superparticle (LSP). If these LSPs
were not in equilibrium at $T_{\rm R}$, the bound $\Omega_{\rm LSP} <
1$ then implies that the reheat temperature must be at least ten
orders of magnitude below the inflaton mass, i.e. the inflaton decay
width must be at least 26 orders of magnitude smaller than $m_\phi$.
This does not seem to be very plausible. As well known, the
requirement $\Omega_{\rm LSP} < 1$ imposes severe constraints on the
parameter space of many supersymmetric models if the LSP was in
equilibrium at $T_R$ \cite{lspconstraint}. Our analysis indicates that
it is quite difficult to evade these constraints by changing the
cosmology. Similarly, we found that stable charged particles can only
be tolerated if they are too heavy to be produced in inflaton decays.

Many of the results presented in this paper are only
semi--quantitative. Unfortunately in most cases significant
improvements can only be made at great effort. For example, a more
accurate treatment of thermalization would require a solution of the
Boltzmann equations in the presence of a non--trivial, but
non--thermal, background of relativistic particles. Once a thermal
plasma has been established, a proper treatment of the slow--down of
primary inflaton decay products would require a careful treatment of
the full momentum dependence of the particle distribution
functions. On the other hand, our estimates of inflaton decay
branching ratios should be quite reliable if $m_\chi > T_{\rm R}$
(which is required for $\chi$ not to have been in thermal equilibrium
at $T_{\rm R}$); even for many--body decays, details of the matrix
elements should change our estimates only be ${\cal O}(1)$
factors. Fortunately we found that this is often also the most
important of the new, non--thermal mechanisms for the production of
massive particles at the end of inflation. We therefore conclude that
cosmological constraints on models with stable or long--lived massive
particles are (much) more severe than had previously been thought.

\section*{Acknowledgements}
We thank S. Davidson, A. Mazumdar and S. Sarkar for valuable
discussions. This work was supported by ``Sonderforschungsbereich 375
f\"ur Astro-Teilchenphysik'' der Deutschen Forschungsgemeinschaft.


\end{document}